\documentclass[a4paper,notitlepage,intlimits,9pt]{amsart}
\numberwithin{equation}{section}
\usepackage{amsthm,amssymb,dsfont,mathrsfs,bbm,bm,xfrac,tabularx}
\usepackage{mathtools}
\usepackage{multirow}
\usepackage{geometry}
\usepackage[export]{adjustbox}
\usepackage[cal=boondoxo]{mathalfa}
\DeclareMathAlphabet\mathrsfso{U}{rsfso}{m}{n}
\DeclareMathAlphabet\mathzapf{T1}{pzc}{mb}{it}
\usepackage[usenames,dvipsnames]{xcolor}
\usepackage[caption=false]{subfig}
\usepackage[english]{babel}
\usepackage{lmodern}
\usepackage{hyperref}
\hypersetup{pdfauthor={Benedetto Caglioti Caracciolo D'Achille Sicuro Sportiello},pdftitle={FSC Assignment},%
            colorlinks, linktocpage=true, pdfstartpage=3, pdfstartview=FitV,%
    breaklinks=true, pdfpagemode=UseNone, pageanchor=true, pdfpagemode=UseOutlines,%
    plainpages=false, bookmarksnumbered, bookmarksopen=true, bookmarksopenlevel=1,%
    hypertexnames=true, pdfhighlight=/O,%
    urlcolor=orange, linkcolor=blue, citecolor=red, %pagecolor=RoyalBlue,%
        }
\usepackage{tikz}
\usetikzlibrary{patterns,decorations,arrows,shapes.geometric,arrows,mindmap,backgrounds,positioning,positioning,fit,backgrounds,positioning,arrows,matrix,calc}
\usetikzlibrary{shapes,backgrounds,mindmap,shadows,shapes.geometric,topaths,snakes,arrows,pgfplots.groupplots,fadings,calc,decorations.markings,positioning,chains,fit}

\DeclareMathOperator{\e}{e}
\newcommand{\dd}{\mathop{}\!\mathrm{d}}
\DeclareMathOperator{\tr}{tr}

\DeclareMathOperator{\Div}{div}
\newcommand{\mediaE}[1]{{\mathbb E\left[{#1}\right]}}
\DeclareMathAlphabet\mathrsfso {U}{rsfso}{m}{n}
\newcommand{\mX}{{\mathrsfso X}}
\newcommand{\mY}{{\mathrsfso Y}}
\newcommand{\mD}{{\mathrsfso D}}

\newcommand{\gab}[1]{{#1}}
\newcommand{\ema}[1]{{#1}}
\newcommand{\spo}[1]{{#1}}

\newcommand{\be}{\begin{equation}}
\newcommand{\ee}{\end{equation}}

\renewcommand{\paragraph}[1]{\noindent {\it #1.}\ }
\newtheorem*{conj}{Conjecture}

\makeatletter
\renewcommand{\email}[2][]{%
  \ifx\emails\@empty\relax\else{\g@addto@macro\emails{,\space}}\fi%
  \@ifnotempty{#1}{\g@addto@macro\emails{\textrm{(#1)}\space}}%
  \g@addto@macro\emails{#2}%
}
\makeatother

\title[Random assignment problems on $2d$ manifolds]{Random assignment problems on $\bm{2d}$ manifolds}
\author[Benedetto]{D.~Benedetto}
\address{Dipartimento di Matematica, Sapienza Universit\`a di Roma, P.le Aldo Moro 5, 00185 Roma, Italy}
\author[Caglioti]{E.~Caglioti}
\address{Dipartimento di Matematica, Sapienza Universit\`a di Roma, P.le Aldo Moro 5, 00185 Roma, Italy}
\author[Caracciolo]{S.~Caracciolo}
\address{Dipartimento di Fisica, Universit\`a di Milano, and INFN, sez.~di Milano,Via Celoria 16, 20100 Milano, Italy}
\author[D'Achille]{M.~D'Achille}
\address{Centre CEA de Saclay, Gif-sur-Yvette, France, CIRB Coll\`ege de France,
11 Place Marcelin Berthelot, 75231 Paris, LI-PaRAD Universit\'e de Versailles Saint-Quentin-en-Yvelines, Versailles and Universit\'e Paris Saclay, France}
\author[Sicuro]{G.~Sicuro}
\address{Department of Mathematics, King's College London, Strand, London WC2R 2LS, United
Kingdom}%\email[G.~Sicuro]{gabriele.sicuro@kcl.ac.uk}
\author[Sportiello]{A.~Sportiello}
\address{ LIPN, and CNRS, Universit\'e Paris 13, Sorbonne Paris Cit\'e, 99 Av.~J.-B.~Cl\'ement, 93430 Villetaneuse, France}
\date{\today}
\allowdisplaybreaks[3]
\begin{document}
\begin{abstract}
We consider the assignment problem between two sets of $N$ random points on a smooth, two-dimensional manifold $\Omega$ of unit area. It is known that the average cost scales as $E_{\Omega}(N)\sim\frac{1}{2\pi}\ln N$ with a correction that is at most of order $\sqrt{\ln N\ln\ln N}$. In this paper, we show that, within the linearization approximation of the field-theoretical formulation of the problem, the first $\Omega$-dependent correction is on the constant term, and can be exactly computed from the spectrum of the Laplace--Beltrami operator on $\Omega$. We perform the explicit calculation of this constant for various families of surfaces, and compare our predictions with extensive numerics.
\end{abstract}
\maketitle
\tableofcontents

%%%%%%%%%%%%%%%%%%%%%%%%%%%%%%%%%%%%%%%%%%%%%%%%%%%%%%%
\section{Introduction}
%%%%%%%%%%%%%%%%%%%%%%%%%%%%%%%%%%%%%%%%%%%%%%%%%%%%%%%

\noindent
The Euclidean assignment problem is a transportation problem between a set $\mX$ of $N$ ``red'' points and a set $\mY$ of $N$ ``blue'' points. Both sets are supposed to be on a given $n$-dimensional Riemannian manifold $\Omega$. A \emph{transportation map} is a bijective map $T\colon\mX\to\mY$, that is, a pairing among the red and blue points. A transportation cost $w_{xy}$ is given for each pair $(x,y)\in\mX\times\mY$, and the cost of the map $T$ is the expression
\begin{equation}
E_{\Omega}[T|\mX,\mY]\coloneqq\sum_{x\in\mX} w_{x\,T(x)}. 
\end{equation}
We will also define $E_{\Omega}[\mX,\mY]$ as the minimum cost among the possible transportation maps. Given a probability distribution for  $\mX$ and a probability distribution for $\mY$, we use $E_{\Omega}(N)$ as a shortcut for $\mathbb{E} [ E_{\Omega}[\mX,\mY]]$.  For definiteness, we will assume in this paper that the points of $\mX$ and $\mY$ are uniformely distributed.

Within this geometric framework, it is natural to choose for $w_{xy}$ a function of the distance $d(x,y)$ between $x\in\mX$ and $y\in\mY$, that is $w_{xy} = f(d(x,y))$.  We expect that, if the function $f$ has some natural (monotonicity, smoothness,\ldots) properties, the large-$N$ behaviour of $E_{\Omega}(N)$ (with the volume of $\Omega$ kept constant) is dominated by the behaviour of $f$ near zero. In turn, this suggests to concentrate only on power-law functions, $f(d)=d^p$ for some $p>0$, as any other detail of the function is either trivially rescaled, or washed out in the limit. The two cases most studied in the literature are $p=1$ and $p=2$, where a number of (different) useful extra features emerge. This paper makes no
exception, and in fact we will only consider here the case $p=2$, that is, we will set once and for all $w_{xy} = d^2(x,y)$.

It is a longstanding question to understand the asymptotic behaviour, for large $N$, of $E_{\Omega}(N)$, and, when $n\geq 2$, the results are very partial for \emph{any} manifold $\Omega$, including the conceptually simplest ones (like the unit hypercube, or the unit hypertorus), and \emph{any} value of $p$, including the special cases $p=1$ and $p=2$. %%% QUID
In particular, in the two-dimensional case for $p=2$ (see \cite{Caracciolo:158,Caracciolo:163,Ambrosio2016,Ambrosio2018}),
it has been proved that\gab{, as long as $\Omega$ has unit volume,} the leading term is $\Omega$-independent:
$$ E_{\Omega}(N)=\frac 1{2\pi} \ln N + o(\ln N).$$
The main goal of the present paper is to show that, at least in the case $n=2$,
$p=2$, the leading behaviour in $N$ of $E_{\Omega}(N)$
which is $\Omega$-dependent is a constant, which can be calculated
\emph{exactly}. More precisely, we are not able to establish a
full perturbative expansion for $E_{\Omega}(N)$, up to corrections $o(1)$,
for \emph{any} $\Omega$. Nonetheless, for \emph{all} pairs $(\Omega, \Omega')$, we predict that  %%%QUID
\be
\label{eq.sketchresult}
E_{\Omega}(N)-E_{\Omega'}(N)=2(K_{\Omega}-K_{\Omega'}) + o(1)
\ee
with
\begin{equation}
  \label{eq:K}
  K_\Omega\coloneqq\lim_{s\to 1}\left[\sum_{i}\frac{1}{\lambda_i^s}-\frac{1}{4\pi}\frac{1}{s-1}\right]
\end{equation}
where $\{\lambda_i\}_{i \geq 1}$ is the set of eigenvalues of the Laplace--Beltrami operator on $\Omega$ that are different from zero (if $\Omega$ is a manifold with a boundary $\partial \Omega$ of perimeter of order 1, it is the set of eigenvalues of the Laplace--Beltrami operator, with Neumann boundary conditions).
That is, all terms in an asymptotic expansion of $E_{\Omega}(N)$ which do not decrease with $N$ must be `universal'.

\ema{One can notice that $K_{\Omega}$ is a regularization of the trace of the Laplace-Beltrami operator. Another equivalent regularization is the so-called Robin mass $R_\Omega,$ see for instance \cite{Okikiolu08,Okikiolu09}. In particular Eq.~\eqref{eq.sketchresult} can be equivalently written as
\be
\label{eq.sketchresultR}
E_{\Omega}(N)-E_{\Omega'}(N)=2(R_{\Omega}-R_{\Omega'}) + o(1)
\ee}
\spo{The definition of $R_{\Omega}$ is postponed to Section~\ref{sec:teoria}, while its relation with $K_\Omega$ is described in Section \ref{ssec.kronetheo}}.

Let us put this result in context, by summarising (part of) the state of the art for this problem.  In contrast with the transportation problem for continuous measures, in this case the candidate optimal transportation maps are just the $N!$ permutations, that is, the possible bijections between two sets of cardinality $N$, and in particular they are a finite set. For one given choice of the $N^2$ weights $w_{xy}$, the computational problem of finding one optimal map $T$,\footnote{Of course, there may be in general more than one optimal map, however, in the random uniform ensemble, the optimal map is almost-surely unique.} and the associated cost $E_{\Omega}[\mX,\mY]$ is a well-studied problem, which turns out to be in the polynomial
class \cite{kuhn1955HungarianMethodAssignment,Jonker1987,lovasz2009matching}. Thus, the associated computational problem can be quickly solved.

\begin{figure}
\subfloat[\label{fig:fibo}]{\includegraphics[width=0.27\columnwidth,valign=c]{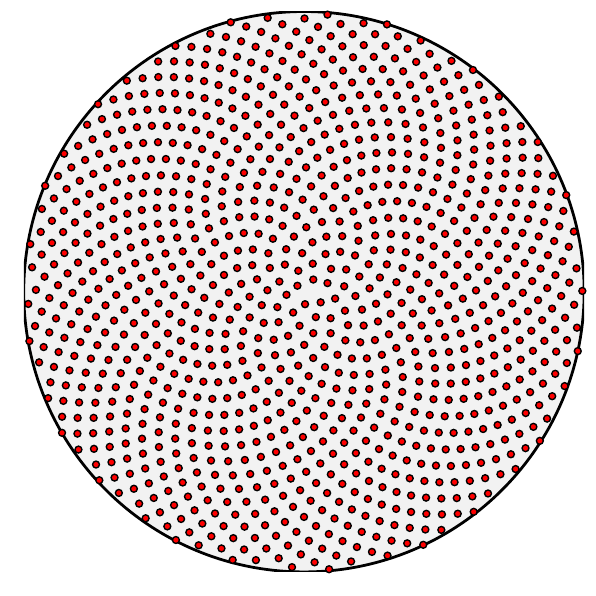}}\hfill
\subfloat[\label{fig:sphere}]{\includegraphics[width=0.27\columnwidth,valign=c]{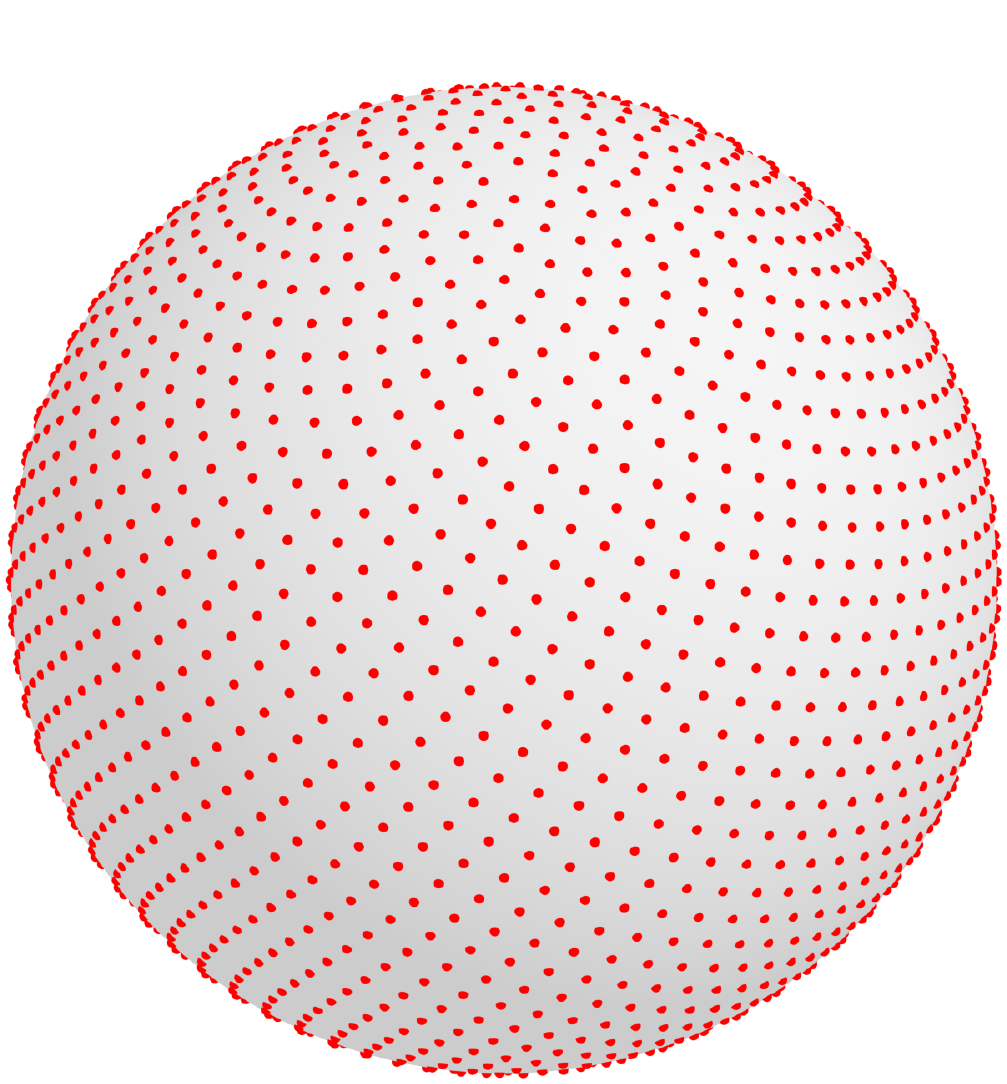}}\hfill
\subfloat[\label{fig:torogrid}]{\includegraphics[width=0.35\columnwidth,valign=c]{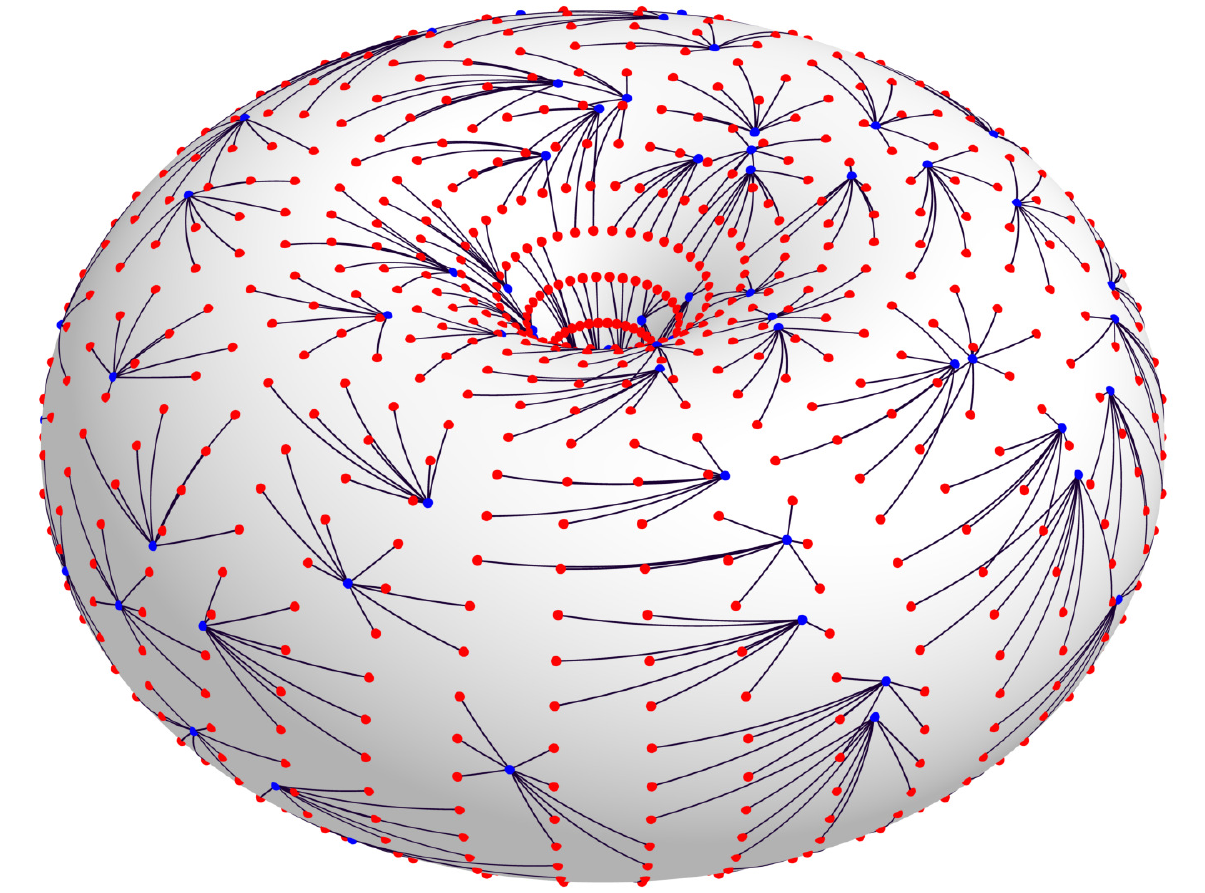}}
\caption{
(a) Fibonacci lattice on the unit disc.
(b) Fibonacci lattice on the unit sphere.
(c) Transportation between a set of $N=144$ random blue points and a square grid of $9N$ red points on the flat torus.}
\end{figure}

 This fact is in striking contrast with the problem in Probability Theory, of understanding the asymptotic of the average cost, on various domains $\Omega$ and statistical ensembles for the red and blue point processes. This random version of the problem has attracted much attention both in Mathematics and Physics. In the Physics community, the interest has come from the analogy with `spin glasses' in Statistical Mechanics, and a seminal contribution was given in the eighties by Orland \cite{Orland1985}, M\'ezard and Parisi \cite{Mezard1985}, that considered the problem ``in infinite dimension'', by introducing the so-called ``random-link'' approximation. This version of the problem was addressed using (non-rigorous) statistical physics techniques such as the replica theory and the cavity method
\cite{Mezard1986a}. Their original results were later put on rigorous ground \cite{Aldous2001,Nair2003,linusson2004}.

The extension to finite-dimension of the random-link results is, however, quite challenging. A first attempt was carried on by M\'ezard and Parisi \cite{Mezard1988,MATCHONELOOP17} that showed how, for $n>2$, the random-link result can be used as a zero-order approximation for the finite-dimensional solution, adding perturbatively a series of corrections. In the same years, a remarkable result was obtained by Ajtai and coworkers \cite{Ajtai} for $n=2$: they proved that, if the problem is considered on the unit square $\Omega\equiv\mathcal{R}\coloneqq[0,1]^2$, then $E_{\Omega}(N)=\mathcal O(\ln N)$.\footnote{More precisely Ajtai \emph{et al.} studied the case $p=1$, but they also sketched how their analysis can be extended to $p$ a positive integer, and predicted the scaling $E_{\Omega}(N)=\mathcal O(N^{1-\frac{p}{2}} (\ln N)^{\frac{p}{2}})$ in this generality.}

Recently, the forementioned result has been refined. In particular, by means of non-rigorous arguments, in Refs.~\cite{Caracciolo:158,Caracciolo:163} it was claimed that, on the unit square $\mathcal{R}$,
\begin{equation}
E_{\mathcal{R}}(N)=\frac{1}{2\pi}\ln N+ 2 c^{\rm PP}_{\mathcal{R}}(N)
\end{equation}
where $c^{\rm PP}_{\mathcal{R}}(N)=o(\ln N)$ (the factor $2$ is for later convenience).  This result has been later rigorously proved by Ambrosio and coworkers \cite{Ambrosio2016} and extended to any $2$-dimensional compact closed manifold $\Omega$ \cite{Ambrosio2018}, where it is shown that 
\begin{equation}
\label{costocorr}
E_{\Omega}(N)=\frac{1}{2\pi}\ln N+ 2 c^{\rm PP}_{\Omega}(N).
\end{equation}
The latter paper also proves rigorous bounds for $c^{\rm PP}_\Omega$, namely that $c^{\rm PP}_\Omega(N)=\mathcal{O}(\sqrt{\ln N\ln\ln N})$.  It has been recently conjectured that Eq.~\eqref{costocorr} holds also in the case of points generated from non-uniform densities \cite{benedetto2019}.

In this paper we further investigate the problem of the estimation of $c_\Omega(N)$. Extending the arguments given in \cite{Caracciolo:163}, we argue that, on a generic two-dimensional manifold of unit area, the correction $c_\Omega$ in Eq.~\eqref{costocorr} can be written as
\begin{equation}\label{correzionintr}
c_\Omega^{\rm PP}(N)=c_*^{\rm PP}(N)+K_\Omega+o(1),
\end{equation}
where $c_*^{\rm PP}(N)$ does not depend on $\Omega$. The index `PP' is to denote that both the red and blue points are sampled with the
`Poisson random process on $\Omega$' (that is, are i.i.d.\ and uniform). Numerical investigations are compatible with the possibility that $c_*^{\rm PP}(N)$ is indeed a constant, and, under this hypothesis, we can give the constant $c_*^{\rm PP}$ the approximate value  
\begin{equation*}
c_*^{\rm PP}=0.29258(2).
\end{equation*}
Analogous claims and results hold for other variants of the problem, most notably when one set of points is still sampled with the Poisson random process, and the other set is either a deterministic regular grid (we investigate here the cases of square (S), triangular (T) or ``Fibonacci'' (F) \cite{Saff1997, RSaff1994} grids), or, in the variant of the problem where $T$ is the transportation
between a discrete and a continuous measure, the uniform measure (U) on $\Omega$. In these three new cases, the factor $2$ in equation
\eqref{costocorr} disappears, and we have the similar structure
\begin{subequations}\label{costoppxp}
\begin{align}\label{costopp}
E_{\Omega}(N)=E_{\Omega}^{\rm PP}(N)&=\frac{\ln N}{2\pi}+ 2 c_*^{\rm PP}(N) + 2 K_{\Omega} +o(1)
\\\label{costoxp}
E_{\Omega}^{\rm \bullet P}(N)=\frac{\ln N}{4\pi}+ c_\Omega^{\rm \bullet P}(N)&\equiv\frac{\ln N}{4\pi}+ c_*^{\rm \bullet P}(N) +K_{\Omega} +o(1)
\qquad
\bullet =%\square,\triangle,
\mathrm{S},\mathrm{T},\mathrm{F},\mathrm{U}.
\end{align}
\end{subequations}
% (We use the Fibonacci grid only in the case of non-flat manifolds).
Let us stress again that the functions $c_*$ are `universal', in the sense that they do not depend on the choice of manifold $\Omega$ (but they do depend on the choice of local randomness, e.g.\ among P, S, T, F, U), while the geometric correction $K_{\Omega}$ depends on the choice of manifold, but is `universal' in a different sense, as it is independent of the choice of local randomness (provided that the extra factor 2 in the P case is taken into account). Just as well as equation \eqref{eq.sketchresult}, such a decomposition is not at all granted \emph{a priori}, and is somewhat surprising.

We also give numerical estimates of the associated values of $c_*$,\footnote{We do not give precise numerical estimates for the Fibonacci grid, as we suppose that, at the size we have investigated, the small variations in the realisation of the Fibonacci grid may affect this constant at an order of magnitude comparable to $c_*^{\rm SP}-c_*^{\rm TP}$, which is numerically rather small.}  under the hypothesis that they are indeed constant, namely
\begin{align}\label{cstarxp}
c_*^{\rm SP}&=0.4156(5)
&
c_*^{\rm TP}&=0.413(2)
&
c_*^{\rm UP}&=0.4038(3).
\end{align}
\bigskip

\noindent The paper is organized as follows. In Section~\ref{sec:defs} we define the random matching problems we are interested in. In Section~\ref{sec:teoria} we present our functional approach for the derivation of the scaling of the optimal cost, including the finite-size corrections given in Eq.~\eqref{correzionintr}. For simplicity, we concentrate only on the Poisson--Poisson case.  In Section~\ref{sec:applica} we apply our theory to different domains, giving an explicit computation of $K_\Omega$ for all of them. In Section~\ref{sec:numerica} we compare our predictions with numerical results obtained solving a large number of instances of the problem on the domains under investigation. Finally, in Section~\ref{sec:conclusion} we give our conclusions.

%%%%%%%%%%%%%%%%%%%%%%%%%%%%%%%%%%%%%%%%%%%%%%%%%%%%%%%
\section{The random assignment problem}
\label{sec:defs}
%%%%%%%%%%%%%%%%%%%%%%%%%%%%%%%%%%%%%%%%%%%%%%%%%%%%%%%

%\subsection{Statement of the problem}
\noindent
Let us consider a connected, two-dimensional smooth Riemannian manifold $\Omega$ having finite volume and, if with a non-empty boundary,
finite perimeter, with metric $g$. Given a system of local coordinates $(x^1,x^2)$ around a point $p\in\Omega$, $g=\sum_{ij}g_{ij}(p)\dd
x^i\otimes\dd x^j$, and given two elements $v,w$ in the tangent bundle in $p$, we will denote by $\langle v,w\rangle_p \coloneqq\sum_{ij}g_{ij}(p) v_iw_j$. For the sake of generality, we will perform our analysis in the slightly more subtle case of $\partial\Omega\neq\emptyset$ (the arguments below can be easily adapted to the case $\partial\Omega=\emptyset$). We will denote $\dd\sigma$ the Riemannian measure for $\Omega$, and we will assume the measure of $\Omega$ to be equal to $1$.\footnote{This is done with no loss of generality, as, of course, under the rescaling $g \to \lambda g$, we have $|\Omega| \to \lambda^2 |\Omega|$ and $E[T|\mX,\mY] \to \lambda^p E[T|\mX,\mY]$, even for a general cost function $f(d)=d^p$, so that the cost for a general surface is trivially deduced from the one for the normalised surface.} Moreover, we will denote by $\delta_p(x)\dd\sigma(x)$ the unit measure concentrated in $p\in\Omega$, that is, given a test function $\varphi(x)$,
\begin{equation*}
%\label{eq.defdelta}
\int_{\Omega} \varphi(x) \delta_p(x)\dd\sigma(x)=\varphi(p).
\end{equation*}
Suppose now that two sets of points are given on $\Omega$, namely a set of $N$ points $\mX\coloneqq\{X_{i}\}_{i=1}^{N} \subset \Omega$,
that we call the red points, and a set of $N$ points $\mY\coloneqq \{Y_{i}\}_{i=1}^{N} \subset \Omega$, that we call the blue points. The
assignment problem consists in assigning each red point to one blue point, in such a way that the resulting map $T$ is a bijection, and a certain total cost function is minimized. As motivated in the introduction, the cost function is the sum of the costs for each pair $(X_i,Y_j)$ such that $T(X_i)=Y_j$, and the cost of a pair $(x,y)$ is the square of the Riemann distance between $x$ and $y$ of the selected pairs. In formulas, we have to find the optimal bijection $T^*\colon\mX\to\mY$ such that
\begin{equation}\label{popt}
T^*\coloneqq \arg\min_{T} E_N[T|\mX,\mY],
\end{equation}
where 
\begin{equation}\label{costo}
E_N[T|\mX,\mY] \coloneqq \sum_{i=1}^N d^2(X_i,T(X_i)),  
\end{equation}
and $d(x,y)$ is the Riemann distance between the points $x$ and $y$, i.e., the infimum of the lengths of the curves that join the two
points. 

Note that each feasible $T$ corresponds to a permutation $\pi$ of $N$ elements, so that $T(X_i)=Y_{\pi(i)}$, and searching for the optimal map is equivalent to searching for the optimal permutation. If we introduce the two atomic measures
\begin{subequations}\label{atomic}
 \begin{align}
 \nu_{\mX}\coloneqq\frac{1}{N} \sum_{i=1}^N \delta_{X_i},\\ 
 \nu_{\mY}\coloneqq\frac{1}{N} \sum_{i=1}^N \delta_{Y_i},
 \end{align}
\end{subequations}
the optimal cost $\min_T E_N[T|\mX,\mY]$ coincides with the $2$-Wasserstein distance (squared) between the two empirical measures
in Eq.~\eqref{atomic}, of which we shall recall here the definition (see e.g.~\cite{Ambrosio2003a}): given two probability
measures $\nu_1$ and $\nu_2$ on $\Omega$, their $2$-Wasserstein distance is
\begin{equation}
\label{eq.JPM}
  W_2^2(\nu_1,\nu_2)\coloneqq \inf_{J\in\Gamma(\nu_1,\nu_2)}\int
  d^2(x,y)\dd J(x,y),
\end{equation}
where the infimum is taken over the set $\Gamma(\nu_1,\nu_2)$ of all the joint probability distributions $J$ with first and second marginal given by $\nu_1$ and $\nu_2$, respectively. It is well known (see for example \cite{Ambrosio2016,ambrosio2006gradient,villani2008,fathi2010}) that, in our setting, the set of the optimal joint probability distributions $J$ is a convex polytope, called Birkhoff polytope, whose extreme points are all and only the permutations $\pi$
%%% QUID
which are optimal within the probability distributions of the form $J(x,y)=\sum_{i} \delta_{X_i}(x) \delta_{Y_{\pi(i)}}(y)$.
Accordingly, the set of optimal maps $T\colon\Omega\to\Omega$ pushing $\nu_1$ to $\nu_2$, i.e., those realising the infimum in the expression
\begin{equation}\label{Wass2T}
  W_2^2(\nu_1,\nu_2)=\inf_{T\colon T_\#\nu_1=\nu_2}\int_{\Omega}d^2(x,T(x))\dd \nu_1(x),
\end{equation}
% are the convex hull of 
coincides with the set of maps of the form $T(X_i)=Y_{\pi(i)}$, with $\pi$ optimal in the sense above. (The situation is much simpler when $\nu_\mX$ is absolutely continuous with respect to the
 Riemannian measure, as in this case the optimal transport map $T$ would be
unique. For a more complete discussion see also \cite[ch.\;9]{villani2008}).

The distance in Eq.~\eqref{Wass2T} corresponds, up to a multiplicative constant, to the cost in Eq.~\eqref{costo} when $\nu_1\equiv\nu_{\mX}$ and $\nu_2\equiv\nu_{\mY}$. Therefore
\begin{equation}
\min_T E_N[T|\mX,\mY]=NW_2^2(\nu_{\mX},\nu_{\mY}).
\end{equation}
In the following, we will consider various statistical ensembles of pairs $(\mX,\mY)$. At this point, many choices are possible. To be
definite, we will always choose $\mX$ and $\mY$ to be independent of each other. We will also choose $\mY$ to be always what we shall call, with abuse of notation, the ``Poisson random process on $\Omega$ of size $N$'', that is, the $Y_j$'s are i.i.d., uniformly chosen on $\Omega$ (w.r.t.\ the measure $\dd\sigma$).\footnote{The `genuine' Poisson random process on $\Omega$ is defined by an intensity, not by a size. The process of intensity $N \dd \sigma$ produces configurations $\mY$ in which the number of points is a Poissonian random variable of average $N$. However, in our context, the large-$N$ convergence of the local properties of the fixed-size Poisson process to the ones of the genuine Poisson process is fast enough to justify our abuse of language.}
\begin{description}
\item[Poisson (P)]Also $\mX$ is given by a Poisson random process on $\Omega$ of size $N$.
\item[Uniform (U)]  Together with the Poisson--Poisson, the most general and interesting
case is the Uniform--Poisson case, in which the cost is
the distance beetwen the Poisson random process and the
uniform measure $\dd \sigma$:
\end{description}
\begin{equation*}
  \label{eq:UP}
  E_{\Omega}^{\rm UP}(N) = N \mathbb{E}\left[W_2^2(\sigma, \nu_{\mY})\right].
\end{equation*}
As a discrete approximation of this case, we can introduce
various  \textit{grid--Poisson assignment problem} (GP),
interesting by themselves:
%% xxxxxxxxxxxxxxxxxxxxxxxxxxxx

%% UNA FRASE DI DUBBIO INTERESSE CHE HO PRODOTTO PER INTRODURRE IL
%% FATTORE 2 FRA PP e GP..........

%% The main feature of the Poisson
%% process is the fact that, for a compact $A \subseteq \Omega$ with $|A|
%% \ll 1$, the number of points of $\mY$ contained in $A$ is a random
%% variable with mean and variance $N|A| (1+ \mathcal{O}(|A|))$, and, for
%% $A$ and $B$ disjoint, the covariance would be of order $N|A||B|$, so
%% that the covariance matrix is almost-diagonal (there would be no
%% corrections at all if we had the `genuine' Poisson point process,
%% however, in our context, the large-$N$ convergence of the local properties of
%% the fixed-size Poisson process to the ones of the genuine Poisson
%% process is fast enough to justify our abuse of language).

%% xxxxxxxxxxxxxxxxxxxxxxxxxxx

%We have various random or deterministic choices for the process generating $\mX$. We will consider in particular
\begin{description}
%\item[Poisson (P)]Also $\mX$ is given by a Poisson random process on $\Omega$ of size $N$.
\item[Square grid (S)]when $\Omega$ is a flat $a \times b$ rectangular 
  domain (possibly up to identification of the boundaries, e.g.\ as in
  a torus), and there exists a value $k$ such that $ka, kb \in
  \mathds{N}$ and $k^2ab=N$, then a natural choice is to fix $\mX$ to
  be the square grid of spacing $1/k$. In the case of a torus, we can
  imagine identifying the horizontal sides of the fundamental rectangular
  region with a shift $s$. In this
  case the grid has no local defects when also $ks \in\mathds{N}$, and
  the modular parameter of the resulting surface is $\tau=(s+i b)/a$,
  so that the set of points in the moduli space which can be realised
  by a grid with cardinality between $N$ and $N +
  N^{\frac{1}{2}+\epsilon}$ becomes dense everywhere in the limit of
  large $N$.
\item[Triangular grid (T)]analogous to the square grid, in the case in
  which $\Omega$ is a flat hexagon (possibly up to identification of
  the boundaries, e.g.\ as in a torus), with sizes $(a,b,c,a,b,c)$ in
  cyclic order. Of course, this includes as special cases the regular
  triangle and hexagon, and the rhombus of angle $\pi/3$. Now we
  require that there exists a value $k$ such that $ka, kb, kc \in
  \mathds{N}$, and $\frac{\sqrt{3}}{2} k^2(ab+bc+ca)=N$, and the
  natural choice is to fix $\mX$ to be the triangular grid of spacing
  $1/k$. In the case of the torus, calling $\omega=\e^{2 \pi i/3}$, the
  associated modular parameter is $\tau=\frac{\omega b + \omega^2 c}
  {a + \omega^2 c}$, so that also in this case the whole moduli space
  can be accessed by increasing $N$.
\item[Fibonacci grid (F)] This is a subtle construction, adapted to
  the case of a sphere, see Fig.~\ref{fig:sphere}, and based on
  stereographic projection from a
  Fibonacci spiral on the plane (from which the name), described in
  \cite{Saff1997, RSaff1994}. The local aspect of this grid around one
  given point is somewhat intermediate between the one of a square and
  of a triangular lattice, with variations depending on the spherical
  coordinates of the point, and on the precise value of $N$. We will
  not enter in the detail of this construction, and the reader is
  referred to the forementioned papers.
\end{description}
% \item[Uniform limit (U)]
In Appendix \ref{app:GPUP} we give more details about the relation between grid-Poisson assignment problems and the UP problem, with an estimation of the convergence rate of the optimal cost in the former to the optimal cost in the latter.
%For any of the grid cases, consider the following procedure: let $h \in \mathds{N}$, and fix one given $\mY$ of size $N$. Call $\mY^{(h)}$ as the set of $hN$ points obtained by taking each point of $\mY$ in $h$ copies, and call $\mX^{(h)}$ as the deterministic grid, for the chosen declination, of size $hN$, see Fig.~\ref{fig:torogrid}. Then the limit $E_{\Omega}[\mY]\coloneqq \lim_{h \to \infty} h^{-1}E_{\Omega}[\mX^{(h)},\mY^{(h)}]$ exists, is independent of the choice of grid recipe, and coincides with the optimal cost as in \eqref{eq.JPM}, with first marginal being the uniform measure, and second marginal given by the measure induced by $\mY$. For further details and a precise estimation of the convergence rate, see .

As anticipated, we are interested in the study of the asymptotic behaviour in $N$ of the average optimal transportation cost, for which
we will adopt the general notation
\begin{equation}
E_{\Omega}(N)\coloneqq \mediaE{\min_T E_N[T|\mX,\mY]}=N\mathbb E\left[W_2^2(\nu_{\mX},\nu_{\mY})\right]
\end{equation}
where the average $\mediaE{\cdot}$ is taken over the pertinent statistical ensemble for the point processes.

%%%%%%%%%%%%%%%%%%%%%%%%%%%%%%%%%%%%%%%%%%%%%%%%%%%%%%%
%%% QUI
\section{Main conjecture}\label{sec:teoria}
As we said above, the study of $E_{\Omega}(N)$ in any dimension $n\neq 1$ or $\infty$ (i.e., in the random-link model) seems rather difficult. A possible approach to
the study of the asymptotic behaviour for large $N$ is based on the fact that, in this limit, $T(x)$ is expected to be `close' to the
identity map (more precisely, from \cite{Ajtai} we expect that $d(X_i,Y_j)=\mathcal O(\sqrt{(\ln N)/ N})$ for pairs of points which are
paired by an optimal matching), and an expansion in this small parameter, at the first non-trivial order, might still capture the relevant features of the solution of the problem. In~\cite{Caracciolo:158,Caracciolo:162,Caracciolo:163} this approach has been applied to the study of the problem when $\Omega$ is the square, or the torus with modular parameter $\tau=i$. The analysis leads to a result whose interpretation requires a regularization that takes into account the finite-$N$ effects and avoid divergences, as we
will see below.

In the approach in \cite{Caracciolo:158,Caracciolo:163,Caracciolo:162}, a close analogy naturally emerges between the evaluation of the average optimal cost in the assignment problem and the evaluation of the electrostatic energy of $2N$ particles, $N$ of each charge sign, pinned in random positions on $\Omega$. This is a result \emph{a posteriori} of the theory, as the obvious analogy just doesn't hold as is (in the electrostatic problem, the energy is the sum of $N(2N-1)$ pair contributions, not just $N$, which scale logarithmically with the distance of the pair, instead that quadratically).  In a sense, the proposed linearization follows the opposite track of the suggestion by Born and Infeld~\cite{Born_Infeld} of a non-linear version of electrodynamics in order to solve the problem of divergencies. Similar ideas have been proposed recently by Brenier for fluid motion~\cite{Brenier2000,Brenier2004}.
\subsection{Linearization}
Let us review the arguments of \cite{Caracciolo:163}, in their natural generalisation to a Riemannian manifold.  We start by introducing, for each map $T\colon\Omega\to\Omega$, the cost
\begin{equation}
\label{cost}
E_N[T|\mX,\mY]=\int_\Omega d^2(x,T(x))\dd \nu_{\mX}(x).
\end{equation}
Note that, at this stage, $T$ is \emph{not} a transportation map. First, because we haven't still imposed the fact that the push-forward of $\nu_\mX$ is $\nu_\mY$, and second, as specific to our transportation problem dealing with atomic measures, the `true' optimal transportation map is only defined on the support of $\nu_\mX$, which is not the whole $\Omega$. We start by solving the first issue. An equivalent formulation of the constraint is that, for any function $\phi\colon\Omega\to\mathds R$, we must have
\begin{equation}
\label{massconserv}
\int_\Omega\phi(T(x))\dd \nu_{\mX}(x)=\int_\Omega\phi(x)\dd \nu_{\mY}(x).
\end{equation}
Again, it would be enough to consider functions $\phi$ with the same support of $\nu_\mY$, a fact of a certain relevance as it implies
that, by expanding $\phi$ over the appropriate basis of functions, we have only $N-1$ independent constraints, instead that infinitely many, as it would be the case if $\nu_{\mY}(x)$ were absolutely continuous with respect to the Lebesgue measure.

The idea is now to write down a Lagrangian that combines the cost expression in Eq.~\eqref{costo} with the condition in Eq.~\eqref{massconserv} as
\begin{equation}\label{eq.lagrFull}
L[T,\phi]\coloneqq \int_\Omega\left[
\frac{1}{2}
d^2(x,T(x))\dd \nu_{\mX}(x)
+\phi(T(x))\dd \nu_{\mX}(x)-\phi(x)\dd \nu_{\mY}(x)\right],
\end{equation}
where $\phi$ plays the role of a Lagrange multiplier. The optimal map $T^*$ satisfies the Euler--Lagrange equations obtained from the Lagrangian above (which turn out to be nonlinear).

% QUI
% It is nowadays common to use field-theoretical methods for studying statistical-mechanics problems \cite{zinn2002,Parisi1998}. When the problem is \emph{at finite temperature}, that is, if we had to evaluate the partition function associated to the Gibbs measure over all possible transportation maps, this would account to study a functional integral, analogous to the ones appearing in Quantum Field Theory (and which are notoriously affected by divergences). Here, however, we are interested in the properties of the optimal map, which are obtained through a suitable \emph{zero-temperature limit}. Also the constraint implemented by the field $\phi$ is `rigid', and thus implemented by a similar zero-temperature limit (although, in general, over a different temperature parameter). As a result, the functional integral exactly reduces to the study of the classical trajectories, which is completely encoded by the Euler--Lagrange equations. So, up to this point, our treatment is exact.%

We shall now use the idea that, for $N\to+\infty$, we expect $T(x)\to x$ for any $x\in\Omega$, due to the fact that the matched pairs become infinitesimally close under the scaling in which $|\Omega|$ is kept fixed. Then, there exists a vector field $\mu(x)$ on $\Omega$ such that, at the leading order
\begin{equation}
\Phi(T(x)) - \Phi(x) = \dd \Phi(\mu)(x)
\end{equation}
and
\begin{equation}
d^2(X_i,T(X_i))=\langle\mu(x),\mu(x)\rangle_{X_i}. 
\end{equation}
The direction of the field is the one of the geodesic curve realising the distance of $T(x)$ from $x$. Pictorially
\begin{center}
 \includegraphics{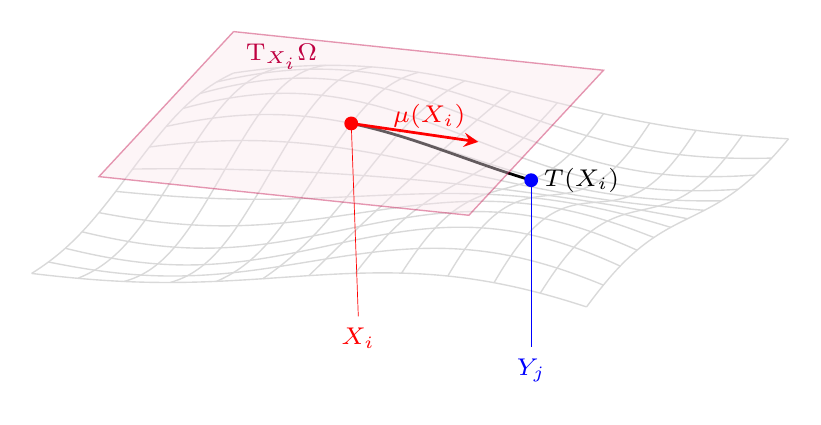}
\end{center}
We shall now introduce
\begin{equation}
 \delta\nu (x)\coloneqq \frac{1}{N}\sum_{i=1}^N\left[\delta_{X_i}(x)-\delta_{Y_i}(x)\right],
\end{equation}
which is another perturbative parameter (when averaging over our statistical ensemble, monomials $\mathbb{E}[\delta\nu (x_1) \delta\nu (x_2) \cdots \delta\nu (x_k)]$ have a definite scaling with $N$, and high powers are suppressed). The Lagrangian is approximated, in this limit, by its quadratic version,
\begin{equation}\label{eq.lagrQuad}
\hat L[\mu,\phi]\coloneqq \int_\Omega 
%\frac{1}{2} 
\left[\frac{1}{2} \langle\mu,\mu\rangle+\langle\mu,\nabla\phi\rangle+\phi\,\delta\nu\right]\dd\sigma,
\end{equation}
where, by definition of gradient, $\dd\phi(\mu)(x)=\langle\mu(x),\nabla\phi(x)\rangle_x$. We have also used the fact that, if the integrand is smooth enough, we can neglect the discrepancy between $\dd \nu_{\mX}$ and $\dd\sigma$ (while still treating more carefully $\delta\nu(x)$). Extremizing the new Lagrangian, and using that if $\partial \Omega \neq\emptyset$ the field $\mu$ is tangent to $\partial \Omega$, we obtain the non-homogeneous linear equations
\begin{subequations}
\begin{align}
\mu&=-\nabla\phi,\\
\Div\mu&=\delta\nu,
\end{align}
\end{subequations}
which are the linearization of the original Euler--Lagrange equations in the fields $\mu$ and $\phi$.  In local coordinates:
\begin{subequations}
\begin{align}
(\nabla\phi(x))_i&=\sum_j g^{ij}\partial_{j}\phi\\
\Div \mu&=\frac{1}{\sqrt{|g|}}\sum_i\partial_{i}\left(\sqrt{|g|}\mu_i\right),
\end{align}
\end{subequations}
where $\partial_i\equiv\partial_{x^i}$, the tensor $g^{ij}$ is the inverse of $g$ and $|g|=\det g$. The two equations imply the Poisson equation 
\begin{equation}\label{eq.Pois}
-\Delta\phi=\delta\nu
\end{equation}
to be solved with Neumann boundary conditions, since the flux of $\mu(x)$ at the boundary is zero.
Here $-\Delta$ is the Laplace--Beltrami operator on $\Omega$, i.e., in local coordinates
\begin{equation}
-\Delta\phi(x)=-\frac{1}{\sqrt{|g|}}\sum_{ij}\partial_i\left(\sqrt{|g|}
g^{ij}\partial_j\phi\right).
\end{equation}

%%%%%%%%%%%%%%%%%%%%%%%%%%%%%%%%%%%%%%%%%%%%%%%%%%%%%%%
\subsection{The divergence of the cost and the problem of regularization}\label{sec:regola}

The functional approach above tells us that $\mu=-\nabla\phi$, where $-\Delta\phi=\delta\nu $. We can use the fact that\footnote{It is only at this point that the linearised theory for the Poisson--Poisson case differs from the theory for the grid--Poisson and the uniform--Poisson cases, as in these cases we would get $1/N$ instead of $2/N$ on the RHS of \eqref{eq.noise}.}
\begin{equation}
\label{eq.noise}
\mathbb E[\delta\nu (x)\delta\nu (y)]=\frac{2}{N}\left(\delta_y(x)-1\right),
\end{equation}
to write down an expression, valid for $N\gg 1$, for a quantity $\epsilon(x)$ that we shall call the \emph{cost density}
\begin{equation}\label{costo1}
\epsilon(x)\coloneqq N\mathbb E\left[|\mu|^2_x\right]=2\int_\Omega\langle\nabla_x G(x,y),\nabla_xG(x,y)\rangle_x\dd\sigma(y),
\end{equation}
so that $E_{\Omega}(N)=\int_\Omega\epsilon(x)\dd\sigma(x)$. Here we have introduced the Green function $G(x,y)$ of $-\Delta$ on the orthogonal complement of the locally constant functions. The Green function is a symmetric function that satisfies the equations
\begin{subequations}\label{Eqgreen}
\begin{align}
-\Delta_y G(x,y)&= \delta_x(y) -1,\\
\left.\partial_{n} G(x,y)\right|_{y\in\partial\Omega}&=0,
\end{align}
where $\left.\partial_{n} G(x,y)\right|_{y\in\partial\Omega}$ is the normal derivative in $x$ with respect to the boundary $\partial\Omega$ of the domain. The equations above identify a unique Green function up to an additive constant: we will fix this constant adopting the convention 
\begin{equation}\label{zeroarea}
\int_\Omega G(x,y)\dd\sigma(x)=0.
\end{equation}
\end{subequations}

The obtained results have, however, a fundamental problem. The quantity in Eq.~\eqref{costo1} is divergent for any $x\in\Omega$. The responsibility for this fact comes from several sources, one of which is having treated the field $\mu(x)$ as a continuous field, instead
that a collection of $N$ vectors, one per each point $X_i \in \mX$. This gives locally, in coordinates on the tangent space in
$X_i$, a field %
\begin{multline}
\mu(x) = -\frac{1}{2\pi N}\frac{x-X_i}{|x-X_i|^2}
+\frac{1}{N}\left.\left[\sum_{j\neq i}\nabla_x G(x,X_j)-\sum_{j}\nabla_x G(x,Y_j)\right]\right|_{\mathclap{\qquad x=X_i}}+\mathcal O\left(|x-X_i|\right)\\
\equiv -\frac{1}{2\pi N}\frac{x-X_i}{|x-X_i|^2} +\hat\mu(x),
\end{multline}
where $\hat \mu(x)$ is such that $\hat \mu(X_i)$ is a finite quantity (here we have used the diagonal expression for the Green function $G$, see below Eq.~\eqref{greenf}). Such an approximation could still be used through a delicate Ces\`aro limit, if we had to perform integrals in which $\mu$ appears linearly. However our cost density is quadratic in these fields, and locally, at a formal level, for $\delta \ll 1$,
\begin{equation}
\frac{1}{\pi\delta^2}\int_{\mathclap{d(x,X_i)<\delta}}
\langle \mu(x), \mu(x)\rangle_x \dd
\sigma(x)
= \frac{1}{2\pi^2\delta^2 N^2}\int_0^{\delta} \frac{\dd r}{r} + 
\langle \hat\mu(X_i), \hat\mu(X_i)\rangle_{X_i}+o(1),
\end{equation}
that is, the appropriate result, which depends on the positions of the points and is finite at finite $N$, is shifted by a fixed but divergent quantity. Yet again, we observe a perfect analogy with 2-dimensional electrostatics, namely with the classical problem of the field self-energy for a distribution of point charges.

Analytically, we see this feature emerging from our result by observing that for $d(x,y)\to 0$, the Green function behaves as \cite{Okikiolu08,Okikiolu09}
\begin{equation}\label{greenf}
 G(x,y)=-\frac{1}{2\pi}\ln d(x,y)+m(y)+\mathcal O(d(x,y)),
\end{equation}
%%%QUID IO: torno indietro, lo giurerei che ln d e' armonica...
%%% QUI
%%where $m(x)=\gamma(x,x),$ and $\gamma(x,y)$ is the regular part of the Green function. 
%%
with a logarithmic divergence.
We perform therefore a regularization of the logarithmic divergence, along the same lines of the classical treatment of
electrostatics. Let us introduce $\Omega_\delta(x)=\Omega\setminus B_\delta(x)$, where $B_\delta(x)=\{y\in\Omega\colon d(x,y)<\delta\}$
is the ball of radius $0<\delta\ll 1$ centered in $x$. We can introduce a regularized expression
\begin{equation}\label{cdelta}
\epsilon_\delta(x)\coloneqq 2\int_{\Omega_\delta}\langle\nabla_x G(x,y),\nabla_xG(x,y)\rangle_x\dd\sigma(y),
\end{equation}
and a corresponding ``regularized cost'' 
\begin{multline}
E_\Omega(\delta)\coloneqq\int_\Omega \epsilon_\delta(x)\dd
\sigma(x)=2\iint_{\mathclap{\substack{\Omega\times\Omega\\d(x,y)>\delta}}}\langle\nabla_x
G(x,y),\nabla_xG(x,y)\rangle_x\dd\sigma(x)\dd\sigma(y)\\
=-2\iint_{\mathclap{\substack{\Omega\times\Omega\\d(x,y)>\delta}}} G(x,y)\Delta_xG(x,y) \dd\sigma(x)\dd\sigma(y)
-2\int_{\Omega}\dd\sigma(y)\int_{\mathclap{\partial
    B_\delta(y)}}G(x,y)\partial_n(x,y)\dd\lambda(x)
\end{multline}
where the second integral runs over the border  $\partial B_{\delta}(y)\coloneqq\{x\in\Omega\colon d(x,y)=\delta\}$ of
$B_{\delta}(y)$, $\dd \lambda(x)$ is the line element of $\partial B_\delta(y)$ in $x$, and $n$ is the outward normal to $B_\delta$. By
Eq.~\eqref{Eqgreen} and Eq.~\eqref{zeroarea} the first integral is infinitesimally small for $\delta\to 0$. Therefore
\begin{equation}
E_\Omega(\delta)=-2\int_{\Omega}\dd\sigma(y)\int_{\mathclap{\partial
    B_\delta(y)}}G(x,y)\partial_nG(x,y)\dd\lambda(x)+\mathcal O(\delta^2|\ln\delta|),
\end{equation}
For $0<\delta\ll 1$, the inner integral can be estimated using the expression in Eq.~\eqref{greenf}, so that
\begin{multline}
 \int_{\mathclap{\partial B_\delta(y)}}G(x,y)\partial_nG(x,y)\dd\lambda(x)=\left[-\frac{\ln\delta}{2\pi}+m(y)+\mathcal O(\delta)\right]\int_{\mathclap{\partial B_\delta(y)}}\partial_nG(x,y)\dd\lambda(x)\\
 =\left[-\frac{\ln\delta}{2\pi}+m(y)+\mathcal O(\delta)\right]\int_{B_\delta(y)}\Delta_x G(y,x)\dd\sigma(x)=\frac{\ln\delta}{2\pi}-m(y)+\mathcal O(\delta).
\end{multline}
We finally get
\begin{equation}\label{costocosto}
E_\Omega(\delta) =-\frac{\ln\delta}{\pi}+2\int_\Omega m(x)\dd\sigma(x)+\mathcal O(\delta).
\end{equation}
%%%% QUI 
The integral of $m(x)$ 
\begin{equation}
  R_\Omega \coloneqq \int_\Omega m(x)\dd\sigma(x),
\end{equation}
is sometimes called \textit{Robin mass} \cite{Okikiolu08,Okikiolu09}. 

Now, we \emph{suppose} that the regularization by the parameter $\delta$ acts in the same way on all geometries.
%%%QUI
%and depends only on the `type' of problem
%{\bf attenzione qui il fattore due c'e' solo per PP} (in our list $\{$P,\ S,\ T,\ F,\ U$\}$), because it is only related to the short-distance regularization of the continuous fields $\mu$ and $\phi$ of the linearised field theory, compared to the exact field theory for the given type of problem. 
\emph{Under this assumption} we can compare two different geometries, $\Omega$ and $\Omega'$, obtaining the following conjecture.
%%% QUI
\begin{conj}
 Let $\Omega$, $\Omega'$ be two regular two-dimensional manifolds, then
 \begin{equation}
\label{costocosto2}
\lim_{N \to \infty}\left( E_{\Omega}(N) - E_{\Omega'}(N) \right)= 2 (R_\Omega-R_{\Omega'}).
\end{equation}
\end{conj}

In other words, the differences of the average cost among different manifolds, in the large-$N$ limit, are expected to be regularization-independent, and, in addition to this, can be expressed in terms \gab{of the Robin's masses of the Laplace--Beltrami Green function on $\Omega$ and $\Omega'$}. The analytic evaluation of these differences will be the main object of our investigation, starting from Section \ref{sec:applica}. The remaining of this section is instead devoted to a further justification of the assumption at the basis of equation \eqref{costocosto2}.

One problem at this point is that our regularization parameter $\delta$ does not have a clear relation with the perturbative parameter $N^{-1}$. In order to better understand what is the
microscopic mechanism beyond the regularization, we observe that equation \eqref{costocosto} can be formally written for $\delta\to 0$ as

\begin{equation}
\label{eq.LBappears}
\lim_{\delta\to 0}E_{\Omega}(\delta)\eqqcolon E_{\Omega} = -2\tr \Delta^{-1}, 
\end{equation}
where the operator $-\Delta^{-1}$ is the inverse Laplace--Beltrami operator on $\Omega$ (with Neumann boundary conditions, if the boundary exists)
\begin{equation}
-\Delta^{-1}\varphi(x)\coloneqq\int_\Omega G(x,y)\varphi(y)\dd \sigma(y).
\end{equation}
As said above, a logarithmic divergence appears for $\delta \to 0$ and both sides of Eq.~\eqref{eq.LBappears} are infinite. By the Weyl law on the asymptotics of the eigenvalue counting function $\mathzapf N_{\Omega}(\lambda)$ for the Laplace--Beltrami operator \cite{ivrii2016} we know that, for a 2-dimensional manifold with unit volume, and under Neumann boundary conditions, the leading behaviour of $\mathzapf N_{\Omega}(\lambda)$ for large $\lambda$ is\footnote{The form of the error term is valid under the assumption that the set of periodic bicharacteristics of $\Omega$ has measure 0 \cite{Duistermaat1975}, while the leading term is valid under no assumption, and was already proven by Weyl, and, by a result of Courant of 1922, we have $\mathzapf N_{\Omega}(\lambda) =\frac{1}{4\pi} \lambda+ \mathcal O(\sqrt{\lambda} \ln \lambda)$ under no assumptions \cite[ch.\;11]{Strauss2008}.}
\be
\mathzapf N_{\Omega}(\lambda) =
\frac{1}{4\pi} 
% \frac{1}{4\pi} 
\left( \lambda + \sqrt{\lambda} |\partial \Omega|
\right)
+ o(\sqrt{\lambda}).
\ee
Furthermore, the eigenfunction $f_{\lambda}$ associated to a given
value of $\lambda$ `looks locally' like a plane wave with wavelength
$1/\sqrt{\lambda}$. This has two consequences at the level of our
approximations when passing from the complete Lagrangian, equation
(\ref{eq.lagrFull}), to its quadratic approximation, equation
(\ref{eq.lagrQuad}).
First, the Taylor expansion of
$\phi(T(x)) = \phi(x+\mu(x))$ around $x$, in the basis of the eigenfunctions
$\{f_{\lambda}\}$, is perturbative in the parameter
$\mathbb{E}(\mu(x)) \sqrt{\lambda}$, which we expect, from
\cite{Ajtai}, to be of order $\sqrt{\lambda \ln N /N}$. Second, if our
basis is orthonormal for the measure $\dd\sigma$,
that is, $(f_{\lambda},f_{\rho})_{\dd\sigma}= \delta_{\lambda \rho}$
(assuming for simplicity of notation that the spectrum is non-degenerate),
under the measure
$\dd \nu_{\mX}$ we get instead
\be
(f_{\lambda},f_{\rho})_{\nu_{\mX}}=
\int_{\Omega} \dd \nu_{\mX}(x)
f_{\lambda}^*(x) f_{\rho}(x)
=
\delta_{\lambda \rho}
+\mathcal{O}\left(
\frac{1}{\sqrt{N}},
\frac{\lambda}{N},
\frac{\rho}{N} \right)
\,.
\ee
The result of this analysis is that, if we decompose our fields in the basis of eigenfunctions of $-\Delta$, we can neglect the corrections coming from the further terms in the Taylor expansion, and the discreteness of the measure, only for those eigenfunctions with $\lambda \lesssim N$ (up to possible factors $(\ln N)^{\gamma}$ in the scaling). Conversely, in the regime $\lambda \gtrsim N$ some unknown mechanism comes into play, and we expect that its effect is to dump the sum $\tr \Delta^{-1}$ appearing in \eqref{eq.LBappears}, possibly at a scale $\lambda \lesssim N$. In \cite{Caracciolo:158}, this unknown dumping mechanism is supposed to be encoded in a cut-off function $F(\lambda/N)$. Now, as this function is related to the local expansion of the fields $\mu$ and $\phi$ at high frequencies, and as 
the relation between eigenvalue $\lambda$ and local wavelength is universal, the function $F(\lambda/N)$ must have one of the two flavours of universality: it shall not depend on the manifold $\Omega$, while in general it must depend on the type of problem (among Poisson, various grids and uniform), that is, in a natural generalisation of the treatment of \cite{Caracciolo:158} to our setting, we should have some unknown functions $F^{\rm \bullet P}(\lambda/N)$, with $\bullet$ being one among P, S, T, F or U, and no dependence on $\Omega$. Going on with our analysis of the PP case (the reasoning can be repeated for all other cases similarly) and using $F(\lambda/N)$ for $F^{\rm PP}(\lambda/N)$ for brevity, within the assumptions of \cite{Caracciolo:158} we should interpret the correspondence \eqref{eq.LBappears} above as
\begin{equation}
\label{Flambda}
E_{\Omega}(N) = 2
\sum_{\lambda\in\Lambda(\Omega)} \frac{F\left(\sfrac{\lambda}{N}\right)}{\lambda} =\int_{0^+}^\infty \frac{F\left(\sfrac{\lambda}{N}\right)}{\lambda}\dd\mathzapf N_\Omega(\lambda),
\end{equation}
where $\Lambda(\Omega)$ is the set of nonzero eigenvalues of the Laplace--Beltrami operator on $\Omega$. Following the analysis already performed in \cite{Caracciolo:158}, this gives
\begin{equation}
\label{costomega}
E_{\Omega}(N)=\frac{1}{2 \pi} \ln N + 2c_{\Omega} + o(1),
\end{equation}
for any domain of unit measure, for some constant $c_\Omega$ depending on the cut-off, which cannot be determined if $F$ is not known. We recall that, as anticipated in the introduction, the leading term in Eq.~\eqref{costomega} is the correct asymptotic cost, as rigorously
proved in Refs.~\cite{Ambrosio2016,Ambrosio2018,Ambrosio2019}, the presence of a logarithm being known since the eighties \cite{Ajtai}.

Note that there is no guarantee that the cut-off function scales exactly as $F(\lambda/N)$, as the mechanism beyond the dumping of the high-wavelength contributions, and the amount of this dumping, are not under control. It may well be, for example, that the function has the form $F\big(\frac{\lambda}{N (\ln N)^\gamma}\big)$, which would give a variant of \eqref{costomega} in which instead of the constant term it will appear an universal term $O(\gamma\ln \ln N).$

All these arguments lead us to reformulate our conjecture as follows.

\begin{conj}[Alternative formulation]
Let $\Omega$ be a regular two-dimensional manifold, then
\begin{equation}
\label{costomega2}
E_{\Omega}(N)=\frac{1}{2 \pi} \ln N + 2 c_*(N) + 2c_{\Omega} + o(1),
\end{equation}
\noindent where $c_*(N)\ema{=o(\ln N)}$ is an universal function not depending on $\Omega$. 

\noindent Moreover, for $\Omega$, $\Omega'$ different regular manifolds, 
\begin{equation}
c_{\Omega} -c_{\Omega'} = R_\Omega-R_{\Omega'} .
\end{equation}

\end{conj}

\noindent {\bf Remark 1.} It is important to remark that, from the simulations made in Section 4, we have evidence that the term $c_*(N)$ can be chosen as a constant, i.e. that formula ~\eqref{costomega} is compatible with our numerical results.
Obviously it is not possibile to deduce that $c_*(N)=c_*$ from numerical simulation only.
Anyway, in case, we would get  ~\eqref{costomega2} and $c_{\Omega} = R_{\Omega}+c_*.$

\vskip.3truecm
\noindent {\bf Remark 2.} 
We notice that starting from Eq.~\eqref{Flambda} and comparing two manifolds, we get the
interesting fact 
\begin{equation}
\label{costocosto2K}
\lim_{N \to \infty}\left( E_{\Omega}(N) - E_{\Omega'}(N) \right)=
2\lim_{N \to \infty}
\int_{0^+}^\infty F\left(\tfrac{\lambda}{N}\right)\frac{\dd\left(\mathzapf N_{\Omega}(\lambda)
-\mathzapf N_{\Omega'}(\lambda)\right)}{\lambda}.\end{equation}
The combination of the two integrals above can be rewritten as
\begin{equation}
\int_{0^+}^\infty F\left(\tfrac{\lambda}{N}\right)\frac{\dd\left(\mathzapf N_{\Omega}(\lambda)
-\mathzapf N_{\Omega'}(\lambda)\right)}{\lambda}=\int_{0^+}^{\infty} \dd \lambda\left(\frac{F(\tfrac{\lambda}{N})}{\lambda^2}-\frac{F'(\tfrac{\lambda}{N})}{N \lambda} \right)
\big(\mathzapf N_{\Omega}(\lambda)-\mathzapf N_{\Omega'}(\lambda)\big).
\end{equation}
The universality of Weyl law implies that the factor $\mathzapf N_{\Omega}(\lambda)-\mathzapf N_{\Omega'}(\lambda)$ grows no faster than $\sqrt{\lambda} \ln \lambda$, so that, even in absence of the function $F$ (that is, in the limit of $N$ large), the integral is convergent at infinity (and near zero is regularised by the spectral gap). This allows us to predict
\begin{equation}
\label{costocosto2Kb}
\lim_{N \to \infty}
\big( E_{\Omega}(N) - E_{\Omega'}(N) \big)
=
2
\int_{0^+}^{\infty} 
\frac{\dd\left(\mathzapf N_{\Omega}(\lambda)
-\mathzapf N_{\Omega'}(\lambda)\right)}{\lambda}.
\end{equation}

\section{Different regularization procedures}
\label{ssec.kronetheo}
The expression \eqref{Flambda} is, annoyingly, a diverging expression depending on $\Omega.$  A way of studying this expression is by introducing a regularization parameter $\epsilon$ for these contributions, and then deducing an evaluation of \eqref{costocosto2Kb} from a singular expansion in $\epsilon$ around zero.

One standard way to perform this programme is the so-called \emph{zeta regularization} \cite{Apostol2012}. Let us introduce the generating function
\begin{equation}
\label{Zfun}
Z_{\Omega}(s)\coloneqq \sum_{\lambda \in \Lambda(\Omega)} \frac{1}{\lambda^{s}},
\end{equation}
which is known to be absolutely convergent for $\Re(s) > 1$, and in this case we recognise our scheme above under the identification $s=1+\epsilon$. Then $-\tr \Delta^{-1}$ can be regularized by looking at $Z_{\Omega}(s)$ near $s=1$ \cite{Osgood1988} 
\begin{equation}
Z_{\Omega}(s) = \frac{1}{4\pi}\frac{1}{s-1} + K_\Omega + \mathcal O(s-1)\, 
\end{equation}
and by removing the pole at $s=1$. That is, in equation \eqref{costocosto2Kb},
\begin{equation}
\label{costocosto2Kc}
\lim_{N \to \infty}\big( E_{\Omega}(N) - E_{\Omega'}(N) \big)
=2\lim_{s \to 1^+}\big(Z_{\Omega}(s)-Z_{\Omega'}(s)\big)=
2 (K_\Omega - K_{\Omega'}).
\end{equation}
For reasons that will appear clearer below, we will call \emph{Kronecker's mass} the constant $K_\Omega$. Despite the fact that there seems to be no reason \emph{a priori} to believe that $K_{\Omega}$ and $R_{\Omega}$ are related, it has been proved by Morpurgo that $R_{\Omega}-K_{\Omega}$ is a universal constant (that is, it does not depend on $\Omega$), given by
\cite{Okikiolu08,Morpurgo2002,Steiner05,Okikiolu08b}
\begin{equation}
\label{diffRK}
R_{\Omega} -K_{\Omega}=-\frac{\gamma_{\mathrm{E}}}{2\pi} + \frac{\ln 2}{2\pi},
\end{equation}
where $\gamma_{\mathrm E}=0.57721\dots$ is the Euler--Mascheroni constant. In particular, this universality result is crucial in checking \emph{a posteriori} that our two predictions \eqref{costocosto2} and \eqref{costocosto2Kc}, obtained by two different analyses, are consistent, and also implies that our Conjecture is equivalent to the statement of E.~\eqref{costocosto2Kb}. The computation of the Kronecker's mass is often easier than the Robin's mass, as we will show below. For a few manifolds $\Omega$, both computations, of $R_{\Omega}$ and $K_{\Omega}$, can be performed with relatively small effort, and we will do this, for pedagogical reasons, in order to illustrate with an example the forementioned general result.

Another way of performing our programme is to consider the regularized sum
\begin{equation}
\label{Wfun}
W_{\Omega}^{(p)}(\epsilon)\coloneqq \sum_{\lambda \in \Lambda(\Omega)}\frac{\exp(-\epsilon \lambda^p)}{\lambda},
\end{equation}
for $p>0$. This corresponds to a specific choice of function $F(\lambda/N)$, provided that $\epsilon$ is identified with $\sfrac{\gamma}{N^p}$, for $\gamma$ some constant. Also in this case we have, universally,
\begin{equation}
\label{Wfunlim}
W_{\Omega}^{(p)}(\epsilon)=-\frac{\ln \epsilon}{4\pi p} + W_{\Omega}^{(p)} + \mathcal{O}(\epsilon),
\end{equation}
and this leads to the prediction
\begin{equation}\label{costocosto2W}
\lim_{N \to \infty}\big( E_{\Omega}(N) - E_{\Omega'}(N) \big)=
2\lim_{\epsilon \to 0^+}\big(W^{(p)}_{\Omega}(\epsilon)-W^{(p)}_{\Omega'}(\epsilon)\big)
=2 (W^{(p)}_\Omega - W^{(p)}_{\Omega'}).
\end{equation}
The analogue of the Morpurgo theorem reads in this case
\begin{equation}
\label{diffWK}
W^{(p)}_{\Omega}-K_{\Omega}=-\frac{\gamma_{\mathrm{E}}}{4p\pi}
\end{equation}
as can be evinced by comparing the two regularizations for the diverging integral $\frac{1}{4\pi}\int_1^{\infty} \frac{\dd x}{x}$ (which, besides the fact that it is an integral rather than a sum, it has all the appropriate asymptotics properties implied by the Weyl
law). Namely, for this choice we have
\be
\frac{1}{4 \pi}\int_1^{\infty} \frac{\dd x}{x^s}
=\frac{1}{4 \pi}\frac{1}{s-1}
\ee
that is, $K=0$, and
\be
\frac{1}{4\pi}\int_1^{\infty} \frac{\dd x}{x}\e^{- \epsilon x^p}
= \frac{\Gamma(0,\epsilon)}{4p\pi}
= \frac{1}{4p\pi}
\left( -\ln \epsilon - \gamma_E + \epsilon + \cdots \right)
\ee
that is, $W^{(p)}=-\frac{\gamma_E}{4p\pi}$.

Another regularization in the same spirit is through the regularized sums
\begin{equation}
\label{WfunR}
W^{\rm sharp}_{\Omega}(\epsilon)\coloneqq 
\sum_{\lambda \in \Lambda(\Omega)}\frac{\theta(\epsilon^{-1}-\lambda)}{\lambda},
\end{equation}
with $\theta(x)$ is the Heaviside step function, which, yet again, corresponds to a specific choice of function $F(\lambda/N)$, provided that $\epsilon$ is identified with $\gamma/N$, for $\gamma$ some constant. Also in this case we have a universal asymptotics
\begin{equation}
\label{WfunlimR}
W^{\rm sharp}_{\Omega}(\epsilon) = -\frac{1}{4 \pi}\ln \epsilon + W^{\rm sharp}_{\Omega} + \mathcal{O}(\epsilon)
\end{equation}
and this leads to the prediction
\begin{equation}
\label{costocosto2WR}
\lim_{N \to \infty}\big( E_{\Omega}(N) - E_{\Omega'}(N) \big)=2\lim_{\epsilon \to 0^+}
\big(W^{\rm sharp}_{\Omega}(\epsilon)-W^{\rm sharp}_{\Omega'}(\epsilon)\big)
=2 (W^{\rm sharp}_\Omega - W^{\rm sharp}_{\Omega'}).
\end{equation}
The analogue of the Morpurgo theorem reads in this case
\begin{equation}
\label{diffWKR}
W^{\rm sharp}_{\Omega}-K_{\Omega}=0,
\end{equation}
as we have
\be
\frac{1}{4 \pi}\int_1^{\epsilon^{-1}} \frac{\dd x}{x}
= -\frac{1}{4 \pi} \ln \epsilon
\ee
that is, $W^{\rm sharp}=0$. 

\section{Examples}
\label{sec:applica}

\noindent To verify our ansatz, we will compute the Kronecker's mass and the Robin's mass for different $\Omega$. We will compare our analytic results with numerical simulations in Section~\ref{sec:numerica}. We will start considering flat manifolds having $g(x)=\mathbb I$, and consider manifolds with uniform curvature starting from Section~\ref{ssec.sfera}.

\subsection{The unit rectangle}
Let us start by considering the problem on the rectangle. We call $\mathcal R(\rho)$ the rectangle $[0,\sqrt{\rho}] \times [0,1/\sqrt{\rho}]$, and we consider the Laplace--Beltrami operator with Neumann boundary conditions. The eigenfunctions of $-\Delta$ on $\mathcal R(\rho)$ are given by 
\begin{equation}
u_{(n,m)}(x,y) = \cos\left( \frac{\pi n x}{\sqrt\rho} \right)\cos\left( \sqrt\rho \pi m y\right),\quad (x,y)\in\mathcal R(\rho),\quad (n,m)\in \mathds N^2\setminus(0,0).
\end{equation}
The corresponding eigenvalues are 
\begin{equation}
\lambda_{(n,m)} = \pi^2 \left(\rho m^2 + \frac{n^2}{\rho} \right),\quad (n,m)\in \mathds N^2\setminus(0,0).
\end{equation}
We proceed computing the Kronecker mass using the regularized function
\begin{equation}
\begin{split}
Z(s)&=\left(\frac{\rho }{\pi^2}\right)^s\sum_{\substack{(n,m)\in \mathds{N}^2\\ n^2+m^2\neq 0}}\frac{1}{(\rho^2m^2 + n^2)^s}
\\
&=\frac{1}{4}\left(\frac{\rho }{\pi^2}\right)^s\sum_{\substack{(n,m)\in \mathds{Z}^2\\ n^2+m^2\neq 0}}\frac{1}{(\rho^2m^2 + n^2)^s}
+\frac{1}{2}\left(\frac{\rho }{\pi^2}\right)^s\left(\sum_{m \geq 1}\frac{1}{(\rho^2m^2)^s}+\sum_{n \geq 1}\frac{1}{(n^2)^s}\right)\\
&=\frac{\zeta_\tau(s)}{4\pi^{2s}}+\frac{\rho^s+\rho^{-s}}{2\pi^{2s}}\sum_{n=1}^\infty\frac{1}{n^{2s}}.
\end{split}
\end{equation}
Here we have adopted $\tau=i\rho$, in compliance with standard notation for modular forms, and have introduced the lattice zeta function $\zeta_\tau(s)$ defined in \ref{app:Kron}. This calculation is readily performed thanks to a remarkable result due to Kronecker (and known as \emph{first limit formula of Kronecker}), reported in \ref{app:Kron}, equation \eqref{Kro1}, and that
we repeat here:
\begin{equation}
\zeta_\tau(s)\coloneqq\sum_{\substack{(n,m)\in
    \mathds{Z}^2\\ n^2+m^2\neq 0}}\frac{[\Im(\tau)]^s}{|n + \tau
  m|^{2s}} =\frac{\pi}{s-1}+2 \pi \left[ \gamma_{\mathrm{E}} - \ln (
  2\sqrt{\Im(\tau)}|\eta(\tau)|^2 )\right] + o(s-1),
\label{Kro1repeat}
\end{equation}
where $\eta(\tau)$ is the Dedekind $\eta$ function. Kronecker's formula allows us to immediately obtain\footnote{We recall here that $\sqrt{\rho}\eta(i\rho)=\eta(\sfrac{i}{\rho})$.}
\begin{equation}
\label{eq.Krect}
K_{\mathcal R}(\rho)=\frac{\gamma_{\mathrm E}}{2 \pi} - \frac{\ln(4\pi^2\rho|\eta(i \rho)|^4)}{4\pi} + 
\frac{1}{12} \left(\rho + \frac{1}{\rho} \right)
\end{equation}
that for $\rho=1$ (unit square) simplifies to
\begin{equation}
K_{\mathcal R}\coloneqq K_{\mathcal R}(1)  =\frac{\gamma_{\mathrm E}}{2\pi}+\frac{\ln(4\pi)}{4\pi}-\frac{\ln\Gamma \left(\sfrac{1}{4}\right)}{\pi} + \frac{1}{6}.
\end{equation}
We will see in the following that the first limit formula of Kronecker will allow us to extract the Kronecker's mass for many types of flat domains: this explains our choice of `Kronecker mass' for denoting $K_\Omega$.

We can give a slightly more compact form to the function in (\ref{eq.Krect}), shifted by its minimal value above:
\begin{equation}
\label{eq.Krectsub}K_{\mathcal R}(\rho) - K_{\mathcal R}(1)
= -\frac{1}{2\pi}  \ln\frac{\eta(i \rho) \eta\left(i\rho^{-1} \right)}{\eta^2(i)}+ \frac{1}{12}\left(\sqrt{\rho} - \frac{1}{\sqrt{\rho}}\right)^2.
\end{equation}
\setcounter{footnote}{0}
We shall make a remark on this expression. In the limit in which the rectangle is very elongated, we get\footnote{The thumb rule in performing these limits is that, for $\rho \to 0, +\infty$,
$$\frac{1}{\pi}\ln \eta(i \alpha \rho) \sim\frac{1}{12} \left(\alpha \rho+\frac{1}{\alpha \rho} \right).$$}
\begin{equation}
\label{eq.1dlimitR}
\lim_{\rho\to \infty}\frac{2K_{\mathcal R}(\rho)}{\rho}=\lim_{\rho\to 0} 2 \rho K_{\mathcal R}(\rho)= \frac{1}{3},
\end{equation}
that is the average cost for the Poisson--Poisson one-dimensional assignment problem on the segment of unit length~\cite{Caracciolo:159}. This is not by accident. Indeed, in any rectangular domain we can evaluate the average energy of the permutation in which the $k$-th red point counting from the left is matched to the $k$-th blue point counting from the left.  This configuration is optimal w.h.p.\ in the limit $\rho \to +\infty$, and would be optimal, at any $\rho$, if the vertical coordinates of all the points were equal. On the other side, a worst case is when all the vertical coordinates of red points are zero, and all the vertical coordinates of blue points are $1/\sqrt{\rho}$, so that, calling $E_{[0,1]}(N)$ the average energy for the 1-dimensional problem on the $[0,1]$ segment, we get
\be
\rho E_{[0,1]}(N)\leq E_{{\mathcal R}(\rho)}(N) \leq \rho E_{[0,1]}(N) + \frac{N}{\rho}
\ee
which, by substituting our scaling ansatz, gives
\be
\rho E_{[0,1]}(N)\leq \frac{1}{2 \pi} \ln N + 2c_*(N) + 2K_{\mathcal R}(\rho)\leq \rho E_{[0,1]}(N) + \frac{N}{\rho}.
\ee
When we take a limit $N \to \infty$, $\rho \to \infty$ on a direction $\rho \gg \sqrt{N}$, we thus get
\be
\frac{1}{3}
\leq 
\frac{2K_{\mathcal R}(\rho)}{\rho}
+
\mathcal{O}\left( \frac{\ln N}{\rho} \right)
\leq 
\frac{1}{3}
+\mathcal{O}\left( \frac{N}{\rho^2} \right),
\ee
which is consistent with (\ref{eq.1dlimitR}).  In the following we will encounter various other domains which allow a consistency check with a 1-dimensional limit.  We will reach similar conclusions, without entering in the details of the estimates, as this is done by minor modifications of the reasonings presented here.

%%%%%%%%%%%%%%%%%%%%%%%%%%%%%%%%%%%%%%%%%%%%%%%%%%%%%%%
\subsection{The flat torus}
\begin{figure}
\subfloat[\label{fig:lattice}]{\includegraphics[width=0.45\columnwidth]{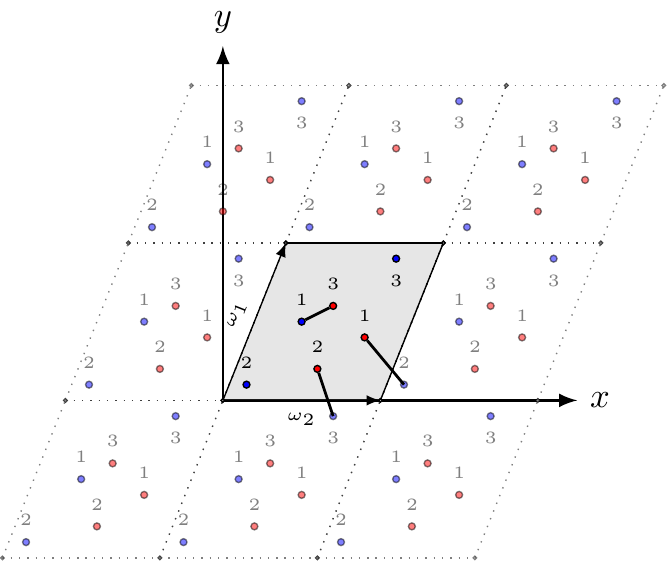}}\hfill
\subfloat[\label{fig:f3}]{\includegraphics[width=0.45\columnwidth]{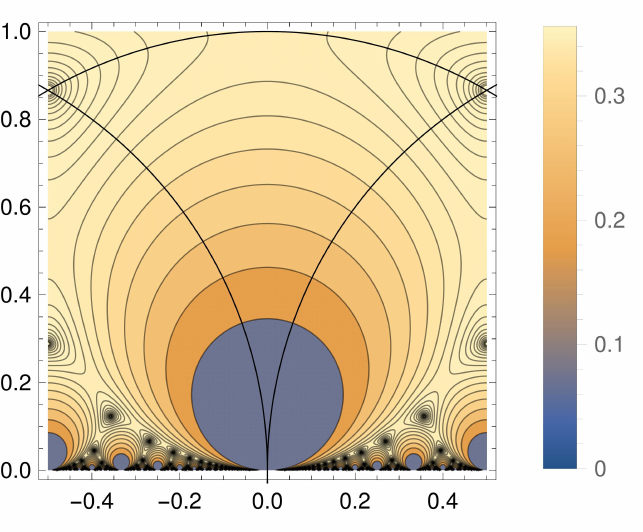}}
\caption{(a) Pictorial representation of an assignment on a torus generated by quotient of $\mathds R^2$ with a periodic lattice, with
  fundamental parallelogram and the corresponding base vectors. (b) Contour plot of $ \Im(\tau) |\eta(\tau)|^4$ in the complex plane
  $\tau$. The triangoloid shape is the canonical fundamental region of the moduli space, given by $|\tau| \leq 1$, $|\tau \pm 1| \geq 1$
  and $\Im(\tau)>0$.}
\end{figure}

We shall now consider the problem on the flat torus $\mathcal T(\tau)$. To describe the corresponding manifold, let us first consider the lattice of points on $\mathds R^2$ 
\begin{equation}
\Lambda= \left\{\underline \omega\ n,\quad n\in\mathds{Z}^2\right\}
\end{equation}
generated by the matrix
\begin{equation}
 \underline\omega\coloneqq \begin{pmatrix}\ell&s\\0&h\end{pmatrix}\quad \ell,h\in\mathds R^+,\ s\in\mathds R,
\end{equation}
corresponding to the base vectors
\begin{equation}
 \omega_1\coloneqq\begin{pmatrix}\ell\\0\end{pmatrix},\quad
 \omega_2\coloneqq\begin{pmatrix}s\\h\end{pmatrix}.
\end{equation}
In such lattice it is possible to define fundamental parallelograms $\mD(\underline \omega)$, containing no further lattice points in its interior or boundary. A fundamental parallelogram is given for example by 
\begin{equation}
\mD(\underline \omega)\coloneqq \{r\in\mathds R^2\colon r=\underline \omega\ x,\quad x\in[0,1)^2\}
\end{equation}
so that $|\mD(\underline \omega)| = \ell h$, see Fig.~\ref{fig:lattice}.  We will also use a shortcut adapted to rectangles,
\begin{equation}
\label{rett}
\mD(\rho)\coloneqq\mD\left(\begin{smallmatrix}\sqrt{\rho} & 0 \\0 & \frac{1}{\sqrt{\rho}}\end{smallmatrix}\right).
\end{equation}
A torus $\mathcal T$ is defined as a quotient between the complex plane and a lattice $\Lambda$, $\mathcal T\coloneqq \mathds R^2/\Lambda$. In other words, each point $x\in\mD$ is identified with the set of points $\{x+\underline\omega\  n,\ n\in\mathds Z^2\}$, the distance between two points in $\mD$ being the minimum distance between the elements of their equivalence classes. It is well known that two matrices $\underline \omega$ and $\underline \omega'$ identify the same lattice $\Lambda$ and the same torus $\mathcal T$ (although not the same fundamental domain $\mD$) if and only if $(\underline \omega)^{-1} \underline \omega' \in \mathrm{SL}(2,\mathds{Z})$. For each $\underline \omega$, we introduce the half-period ratio
\begin{equation}
\tau\coloneqq \frac{s+ih}{\ell}\in\mathds C.
\end{equation} 
Given a lattice $\Lambda$ generated by $\underline\omega$, it is possible to associate to it a dual lattice $\Lambda^*$ generated by
$\underline\omega^*$, such that $\underline\omega^*\ \underline\omega^T=\mathbb I$, identity matrix, i.e.,
\begin{equation}
\underline\omega^*\coloneqq \frac{1}{h \ell}\begin{pmatrix}h&0\\-s&\ell\end{pmatrix}.
\end{equation}
Each torus $\mathcal T=\mathds R^2/\Lambda$ is then naturally associated to a dual torus given by $\mathcal T^*\coloneqq\mathds R^2/\Lambda^*$. 

In the following, we will restrict, without loss of generality, to the case in which the fundamental parallelograms have unit area, choosing
\begin{equation}
\label{eq.omerhosig}
\underline\omega=\underline\omega(\tau)=
\frac{1}{\sqrt\rho}\begin{pmatrix}1&\sigma\\0&\rho\end{pmatrix},
\qquad
\tau\coloneqq\sigma+i\rho,
\end{equation}
such that $\rho\in\mathds R^+$ and $\sigma\in\mathds R$, and we will denote the corresponding torus by $\mathcal T(\tau)$, where $\tau\coloneqq\sigma+i\rho$ is the half-period ratio. 

\paragraph{The Kronecker's mass}Due to the periodicity conditions, the eigenfunctions of $-\Delta$ on $\mathcal T(\tau)$ have the form 
\begin{equation}
u_{k^*}(x) = \exp(2 \pi i\; k^*\cdot x)
\end{equation}
for all $k^*=\underline\omega^*\  k\in\Lambda^*$, $k=\binom{n}{-m}\in\mathds Z^2$. The corresponding eigenvalue is 
\begin{equation}
\lambda_{(n,m)}=|2 \pi k^*|^2
=(2\pi)^2 \frac{|n+\tau m|^2}{\rho}
=(2\pi)^2 \frac{|n+\tau m|^2}{\Im(\tau)}.
\end{equation}
We can compute now the Kronecker mass using the regularized function
\begin{equation}
Z(s)=\sum_{k^*}\frac{1}{|2\pi k^*|^{2s}}=\frac{1}{(2\pi)^{2s}}\sum_{\substack{(m,n)\in \mathds{Z}^2\\ n^2+m^2\neq 0}}\frac{[\Im(\tau)]^s}{|n+\tau m|^{2s}}
\end{equation}
and removing the pole in $s\to 1$, as discussed in Section~\ref{sec:teoria}. This calculation is readily performed, again thanks to the first limit formula of Kronecker, equation \eqref{Kro1}, which allows us to immediately obtain
\begin{equation}
K_{\mathcal T}(\tau)\coloneqq\frac{\gamma_{\mathrm E}}{2 \pi} -\frac{1}{4\pi} \ln \left(16\pi^2\Im(\tau) |\eta(\tau)|^4
\right). \label{CKtau}
\end{equation}
In Fig.~\ref{fig:f3} we present a contour-plot of the related expression $\Im(\tau) |\eta(\tau)|^4$ in the complex plane $\tau$,
confined to the canonical fundamental region of the moduli space. In particular, this function diverges for $\tau \to 0$ and has minimum at $\tau=\exp(i \pi (\frac{1}{2} \pm \frac{1}{6}))$. This implies that, among all unit tori equipped with the flat metric, the ``hexagonal'' one, that is the one for which $\tau$ is a sixth root of unity, is the one in which the average cost of the Euclidean Random Assignment Problem is minimised. More strikingly, as deduced from results in~\cite{Okikiolu08} which are in turn based on the results in~\cite{Osgood1988}, the hexagonal torus is minimal also among unit surfaces with non-uniform metric.

%%%%%%%%%%%%%%%%%%%%%%%%%%%%%%%%%%%%%%%%%%%%%%%%%%%%%%%
\paragraph{Example: the rectangular and rhomboidal tori} We shall call ``rectangular torus'' a torus in which the fundamental parallelogram is a rectangle. This case corresponds to $\tau=i\rho$,
with $\rho>0$ real. Our formula specialises to 
\begin{equation}
K_{\mathcal T}(i\rho)=\frac{\gamma_{\mathrm E}-\ln(4\pi\sqrt\rho)}{2\pi} - \frac{1}{\pi} \ln |\eta(i \rho)|,
\end{equation}
which is invariant under the map $\rho\mapsto \rho^{-1}$, as it should. In the region $\rho\in(0,1]$ the lowest value is achieved at $\rho=1$ (see also Fig.~\ref{fig:f3}), where
\begin{equation}
K_{\mathcal T}\coloneqq K_{\mathcal T}(i)=\frac{\gamma_{\mathrm E}}{2\pi}+ \frac{\ln\pi}{4 \pi}  - \frac{1}{\pi}
\ln\Gamma\left(\sfrac{1}{4}\right).
\end{equation}
We can give a slightly more compact form to this function, shifted by its minimal value:
\begin{equation}
\label{eq.KrectT}
K_{\mathcal T}(i\rho) - K_{\mathcal T}(i)
= -\frac{\ln\rho}{4\pi} -\frac{1}{\pi}  \ln\frac{\eta(i \rho) }{\eta(i)}\\
=  -\frac{1}{2\pi}  \ln\frac{\eta(i \rho) \eta\left(i\rho^{-1} \right)}{\eta^2(i)}.
\end{equation}
Similarly, we shall call ``rhomboidal torus'' a torus in which the fundamental parallelogram is a rhombus. This case corresponds to 
$\tau =\e^{i \theta}$, with $0< \theta \leq \sfrac{\pi}{2}$, and our formula specialises to
\begin{equation}
K_{\mathcal T}(\e^{i\theta})= \frac{\gamma_{\mathrm E}-\ln(4\pi)}{2\pi}- \frac{1}{4\pi} \ln\sin \theta  - \frac{1}{\pi} \ln |\eta(\e^{i \theta})|,
\end{equation}
that is, again shifting by the value for the standard torus,
\begin{equation}
K_{\mathcal T}(\e^{i\theta})-K_{\mathcal T}(i)
=- \frac{1}{4\pi} \ln\sin \theta  - \frac{1}{\pi} 
\ln\frac{2 \pi^{\sfrac{3}{4}}|\eta(\e^{i \theta})|}{\Gamma(\sfrac{1}{4})}.
\end{equation}
As was the case for the rectangle, the expression in equation \eqref{eq.KrectT}, in the limit in which the torus is very ``thin and
long'', becomes 
\begin{equation}
\label{eq.1dlimitT}
\lim_{\rho\to \infty}\frac{2K_{\mathcal T}(i\rho)
}{\rho}
=\lim_{\rho\to 0}
2 \rho K_{\mathcal T}(i\rho)
=
\frac{1}{6}.
\end{equation}
This happens to be the average cost for the Poisson--Poisson
one-dimensional assignment problem on the circle of unit
length~\cite{Caracciolo:159}, as was to be expected, by a reasoning
analogous to the one presented for the case of the rectangle.

\begin{figure}
\subfloat[\label{fig:ct}]{\includegraphics[width=0.4\columnwidth]{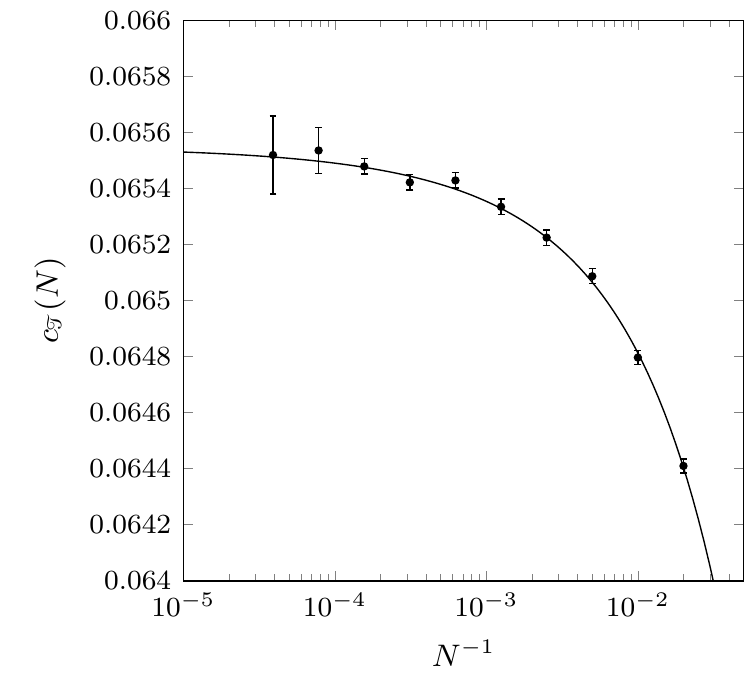}}
\subfloat[\label{fig:qtb}]{\includegraphics[width=0.4\columnwidth]{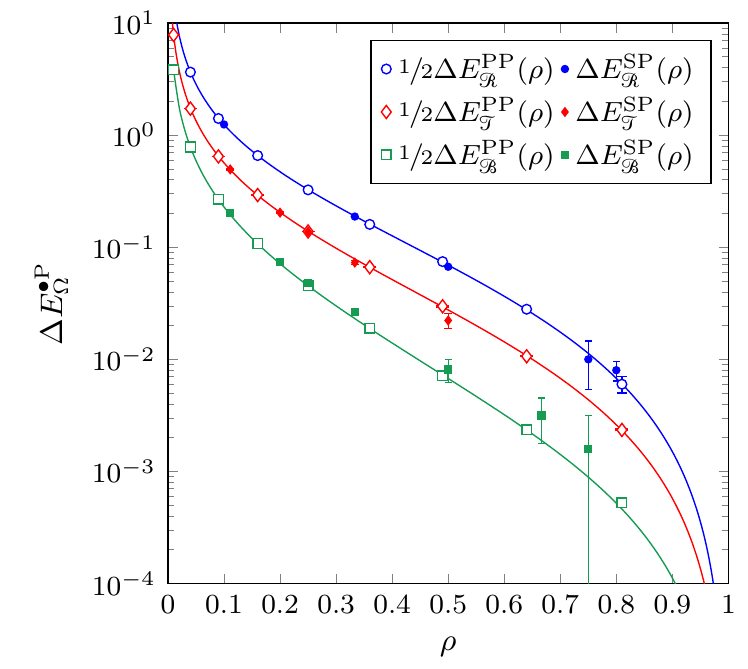}}
\caption{(a) Numerical estimation for $c_{\mathcal T}(N)$ for different values of $N$. Each data point is obtained averaging the optimal cost over at least $10^6$ instances and then removing the leading $\frac{1}{2\pi}\ln N$ term. The fit is obtained using a quadratic function in $\sfrac{1}{\sqrt N}$. (b) Difference of average optimal costs for the assignment on the rectangle $\mathcal R(\rho)$, on the torus $\mathcal T(i\rho)$ and on the Boy surface $\mathcal B(\rho)$ with the corresponding costs for $\rho=1$. The numerical results, represented by the dots, are compared with the analytical prediction obtained from Kronecker's masses.}
\end{figure}

%%%%%%%%%%%%%%%%%%%%%%%%%%%%%%%%%%%%%%%%%%%%%%%%%%%%%%%
\paragraph{The Robin mass} Let us now evaluate, for the generic flat torus $\mathcal T(\tau)$, the Robin mass $R_{\mathcal T}(\tau)$.  Calling $z=z(x,y)= (x_1 -y_1) + i(x_2 - y_2)$, the Green's function on the torus is given in this case
by \cite{Lin2010}
\begin{equation}
G(x,y) =-\frac{1}{2\pi} \ln\frac{\theta_1\left( \sqrt{\Im(\tau)} \, z ; \tau \right)}{\eta(\tau)} +\frac{\Im(\tau) \, (\Im(z))^2}{2}
\end{equation}
where $\theta_1(z; \tau)$ is an elliptic $\theta$ function. The Robin mass is obtained from
\begin{equation}
R_{\mathcal T}(\tau)\coloneqq -\lim_{z\to 0} \left[\frac{1}{2\pi} \ln\left|\frac{\theta_1\left(\sqrt{\Im(\tau)} z; \tau\right)}{\eta(\tau
    )} \right|-\frac{\ln |z| }{2\pi} \right]   =  -\frac{1}{4\pi} \ln\left[ 4\pi^2\Im(\tau)|\eta(\tau)|^4\right].
\end{equation}
It is immediately seen that, in agreement with the Morpurgo theorem, equation \eqref{diffRK} is satisfied.

%%%%%%%%%%%%%%%%%%%%%%%%%%%%%%%%%%%%%%%%%%%%%%%%%%%%%%%
\subsection{Other boundary conditions on the unit rectangle}
\label{sec:klein}
The unit rectangle and the rectangular torus are obtained starting from the fundamental domain $\mD(\rho)$ in equation \eqref{rett}, and
assuming respectively open (i.e., Neumann for the field $\phi$) and periodic boundary conditions. Other choices of boundary conditions are possible, which correspond to other classical surfaces, with or without boundary. Each choice leads to a different spectrum of the Laplacian, which in turn implies a different Kronecker mass (and, according to our theory, a different finite-size correction to the
optimal cost of the assignment problem).

\subsubsection{The cylinder}
Let us consider the domain $\mD(\rho)$ and let us take periodic boundary conditions in the horizontal direction (i.e., the side of length $\sqrt\rho$) and Neumann boundary conditions in the vertical direction (i.e., the side of length $\sfrac{1}{\sqrt\rho}$), see
Fig.~\ref{fig:cymo}. The resulting surface is a cylinder, that we shall denote by $\mathcal C(\rho)$. The eigenfunctions of $-\Delta$ are the set of functions
\begin{equation}
u_{(m,n)}(x,y) =\exp\left(\frac{2i \pi m x}{\sqrt\rho}\right)\cos\left(\pi\sqrt\rho n y\right),\qquad m\in \mathds Z,\ n\in \mathds{N}.
\end{equation}
The corresponding eigenvalues are therefore
\begin{equation}
\lambda_{(m,n)} = \pi^2 \left(4 \frac{m^2}{\rho} + \rho n^2 \right),
\qquad m\in \mathds Z,\ n\in \mathds{N}.
\end{equation}
Repeating the same type of calculations performed for the rectangle (that is, expressing the regularised sum as a combination of
$\zeta_{\tau}(s)$ (for some $\tau$'s) and $\zeta(2s)$), we obtain
\begin{equation}
K_{\mathcal C}(\rho) =\frac{\gamma_{\text{E}}}{2 \pi} - \frac{\ln (16 \pi^2 \rho)}{4\pi}  -  \frac{1}{\pi} \ln \eta(2i  \rho)  +
\frac{1}{24\rho}
\end{equation}
so that
\begin{equation}
K_{\mathcal C}\coloneqq K_{\mathcal C}(1)=\frac{\gamma_{\text{E}}}{2\pi} + \frac{3 \ln 2}{8 \pi}  + \frac{\ln \pi}{4 \pi}  - \frac{\ln \Gamma \left(\sfrac{1}{4} \right)}{\pi}  + \frac{1}{24}.
\end{equation}
We also remark that
\begin{equation}
\lim_{\rho \to \infty} \frac{2K_{\mathcal C}(\rho)}{\rho} = \frac{1}{3},
\end{equation}
which is the cost density for the one-dimensional assignment problem with open boundary conditions (i.e., on the unit segment), while
\begin{equation}
\lim_{\rho \to 0}2 \rho K_{\mathcal C}(\rho)= \frac{1}{6},
\end{equation}
which is the density of cost for the one-dimensional assignment problem with periodic boundary conditions (i.e., on the unit circle),
again, as was to be expected. The nontrivial solution of the equation $K_{\mathcal C}(\rho)=K_{\mathcal C}$ is $\rho=0.625352\dots$, while the minimum value of the mass occurs for $\rho=0.793439\dots$

Remark that the constants above do \emph{not} appear in the study of \cite{Osgood1988}, because in our context, in presence of a boundary, we should impose Neumann boundary conditions (while the authors of \cite{Osgood1988} only analyse the case of Dirichlet boundary conditions).

%%%%%%%%%%%%%%%%%%%%%%%%%%%%%%%%%%%%%%%%%%%%%%%%%%%%%%%
\subsubsection{The M\"{o}bius strip}
\begin{figure*}
\subfloat[\label{fig:cymo}Cylinder.]{\makebox[2cm][c]{\includegraphics[height=0.3\textwidth]{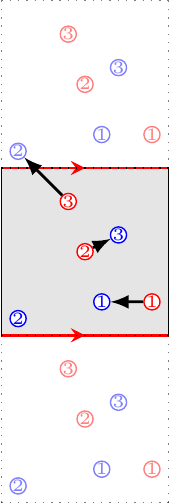}}}
\subfloat[\label{fig:cymo2}M\"{o}bius strip.]{\makebox[2.4cm][c]{\includegraphics[height=0.3\textwidth]{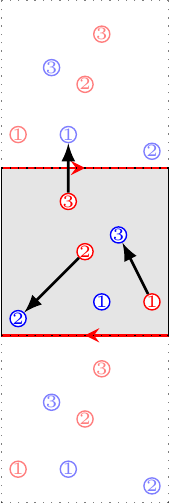}}}
\subfloat[\label{fig:kl}Klein bottle.]{\includegraphics[height=0.3\textwidth]{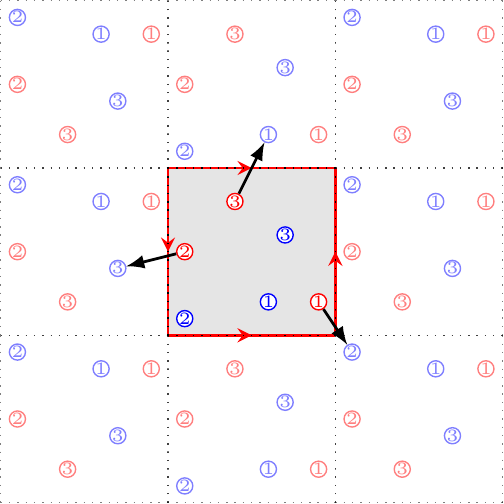}}\hfill
\subfloat[\label{fig:bo}Boy surface.]{\includegraphics[height=0.3\textwidth]{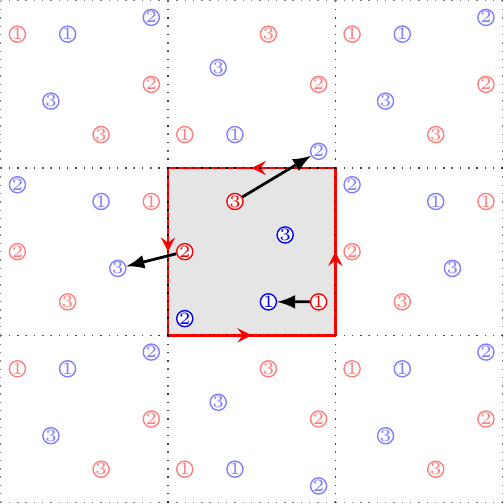}}
\caption{Pictorial representation of the different boundary conditions considered in Section~\ref{sec:klein}, with an example of assignment at $N=3$ for each case. To obtain the corresponding surface, the red edges have to be considered joined in such a way that the directions of the arrows match.}
\end{figure*}

Starting again from the rectangle $\mD(\rho)$, we can identify each point $(x,y)\in\mD(\rho)$ with all its images in $\mathds{R} \times [0,\sfrac{1}{\sqrt\rho}]$ generated by the map $(x,y)\to (x+\sqrt\rho,\sfrac{1}{\sqrt\rho}-y)$. That is, if we see the surface as the fundamental rectangular domain $\mD(\rho)$, we impose open boundary conditions along the horizontal direction, and identify the two vertical sides after a `twist' (that is, the top part on the left is glued to the bottom part on the right). The obtained domain $\mathcal M(\rho)$ is called the \emph{M\"{o}bius strip}, see Fig.~\ref{fig:cymo2}. The eigenfunctions of $-\Delta$ are 
\begin{equation}
u_{(m,n)}(x,y) =\exp\left(\frac{i \pi m
  x}{\sqrt\rho}\right)\cos\left(\pi\sqrt\rho n y\right),\qquad  m\in \mathds{Z},\ n\in \mathds{N},\ m+n \textrm{~even.}
\end{equation}
The parity constraint implements the twisted identification of the strip. The corresponding eigenvalues are
\begin{equation}
\lambda_{(m,n)} = \pi^2 \left(\rho^{-1}m^2 + \rho n^2\right).
\end{equation}
Repeating now the usual arguments we get
\begin{equation}
K_{\mathcal M}(\rho)=\frac{\gamma_{\text E}}{2 \pi} - \frac{\ln (4 \pi^2 \rho)}{4\pi}- \frac{1}{\pi}\ln \frac{\eta^3(i \rho)}{\eta(2i \rho)\eta\left(i\frac{\rho}{2}\right)}+ \frac{1}{24 \rho},
\end{equation}
so that
\begin{equation}
K_{\mathcal M}\coloneqq K_{\mathcal M}(1)= \frac{\gamma_{\text E}}{2\pi} + \frac{\ln(2\pi)}{4 \pi} - \frac{\ln\Gamma \left(\sfrac{1}{4}\right)}{\pi} + \frac{1}{24}.
\end{equation}
We also remark that
\begin{equation}
\lim_{\rho \to \infty} \frac{2K_{\mathcal M}(\rho)}{\rho} = \frac{1}{12}
\end{equation}
which is the average cost for the problem on the segment of length $\frac{1}{2}$ (and the fact that the length is not 1 is related to the
fact that the twisted boundary conditions are effectively `folding in two' the segment), while
\begin{equation}
\lim_{\rho \to 0} 2 \rho K_{\mathcal M}(\rho)= \frac{1}{6},
\end{equation}
which is the cost for the problem on the unit circle. The non-trivial solution of the equation $K_{\mathcal M}(\rho) = K_{\mathcal M}$ is found for $\rho=4.1861\dots$, whereas the minimum is achieved at $ \rho=2.30422\dots$

%%%%%%%%%%%%%%%%%%%%%%%%%%%%%%%%%%%%%%%%%%%%%%%%%%%%%%%
\subsubsection{The Klein bottle}
If we identify both pairs of opposite sides of the rectangle, one pair (say, the horizontal sides) in the ordinary way, and the other pair in the twisted way as in the M\"{o}bius strip, we obtain the \emph{Klein bottle} $\mathcal K(\rho)$, see
Fig.~\ref{fig:kl}. The eigenfunctions of $-\Delta$ are in this case
\begin{equation}
u_{(m,n)}(x,y)=\e^{\frac{\pi i m}{\sqrt\rho} x}\cos \left(2 \pi n\sqrt\rho y\right), \qquad m\in \mathds{Z},\ n\in \mathds{N},\ m+n\text{\ even}
\end{equation}
and
\begin{equation}
v_{(m,n)}(x,y)=\e^{\frac{2m+1}{\sqrt\rho}\pi i x} \sin \left(2 \pi n\sqrt\rho y\right),\qquad m\in \mathds{Z},\ n\in \mathds{N}^+.
\end{equation}
Proceeding as above, one can obtain
\begin{equation}
K_{\mathcal K}(\rho)=\frac{\gamma_{\text E}}{2 \pi} - \frac{\ln(4 \pi^2 \rho)}{4\pi} - \frac{1}{\pi}\ln\eta\left(i \frac{\rho}{2}\right)  - \frac{\zeta(2)}{2 \pi^2 \rho}
\end{equation}
so that, in particular,
\begin{equation}
K_{\mathcal K}\coloneqq K_{\mathcal K}(1)=\frac{\gamma_E}{2 \pi} +\frac{7}{8 \pi} \ln 2  + \frac{1}{4 \pi} \ln \pi - \frac{\ln\Gamma
  \left(\sfrac{1}{4} \right)}{\pi} - \frac{1}{12}.
\end{equation}
We also remark that both one-dimensional limits coincide with the corresponding constructions for the M\"obius strip, and indeed the
limits of the analytical expressions are the same, as
\begin{equation}
\lim_{\rho \to \infty} \frac{2K_{\mathcal K}(\rho)}{\rho} = \frac{1}{12}, 
\end{equation}
while
\begin{equation}
\lim_{\rho \to 0} 2 \rho K_{\mathcal K}(\rho)= \frac{1}{6}.
\end{equation}
Here $K_{\mathcal K}(\rho) = K_{\mathcal K}$ for $\rho = 1.09673\dots$, whereas the minimum is obtained at $\rho = 1.04689\dots$.

%%%%%%%%%%%%%%%%%%%%%%%%%%%%%%%%%%%%%%%%%%%%%%%%%%%%%%%
\subsubsection{The Boy surface}
As a final example, let us take twisted boundary conditions for both pairs of opposite sides of $\mD(\rho)$. In this way we obtain the
so-called \emph{Boy surface} $\mathcal B(\rho)$, see Fig.~\ref{fig:bo}. The eigenfunctions of $-\Delta$ are 
\begin{subequations}
\begin{gather}
u_{(m,n)}(x,y)=\cos \left(\frac{ \pi m}{\sqrt\rho} x\right)\cos
\left(\pi n\sqrt\rho y\right),
\qquad
m,n\in \mathds{N},\ m+n\text{\ even,}
\\
v_{(m,n)}(x,y)=\sin \left(\frac{ \pi m}{\sqrt\rho} x\right)\cos
\left(\pi n\sqrt\rho y\right),
\qquad m,n\in \mathds{N},\ m+n\text{\ odd.}
\end{gather}
\end{subequations}
The calculation proceeds as in the other cases, giving
\begin{equation}
K_{\mathcal B}(\rho)=\frac{\gamma_{\text E}}{2 \pi} - \frac{\ln (4 \pi^2 \rho)}{4\pi}   - \frac{\ln \eta\left(i \rho\right)}{\pi}
-\frac{1}{24}\left(  \rho+\frac{1}{\rho} \right)
\end{equation}
which is symmetric for $\rho \leftrightarrow \frac{1}{\rho}$, as it must be. In particular
\begin{equation}
K_{\mathcal B}\coloneqq K_{\mathcal B}(1)=\frac{\gamma_{\text E}}{2 \pi} + \frac{3}{8 \pi} \ln 2  + \frac{1}{4 \pi} \ln \pi - \frac{\ln\Gamma\left(\sfrac{1}{4} \right)}{\pi}  - \frac{1}{12}.
\end{equation}
Now, both one-dimensional limits produce a domain corresponding to the segment of length $\frac{1}{2}$, and indeed
\begin{equation}
\lim_{\rho \to \infty}\frac{K_{\mathcal B}(\rho)}{\rho}=\lim_{\rho \to 0} 2 \rho K_{\mathcal B}(\rho)
= \frac{1}{12}.
\end{equation}

\begin{figure}\includegraphics[width=\columnwidth]{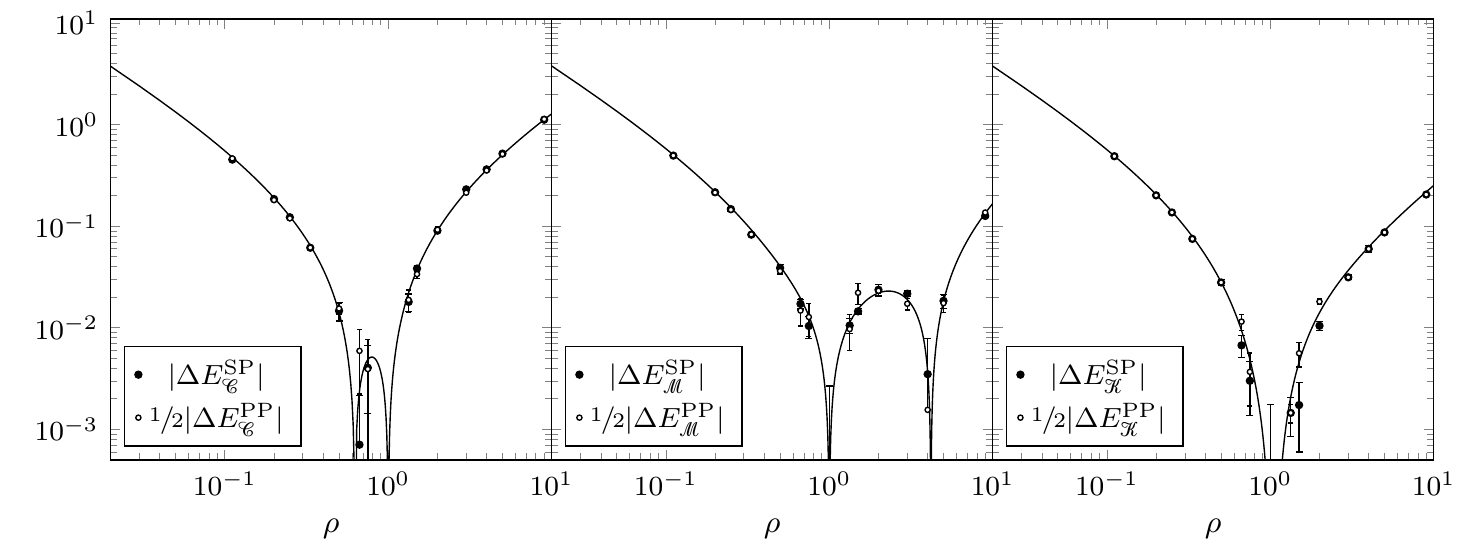} 
\caption{Absolute difference of average optimal costs for the assignment on the cylinder $\mathcal C(\rho)$, on the M\"{o}bius
  strip $\mathcal M(\rho)$ and on the Klein bottle $\mathcal K(\rho)$ with the corresponding costs for $\rho=1$. The numerical results,
  represented by the dots, are compared with the analytical prediction obtained from Kronecker's masses.\label{fig:cmk}}
\end{figure}

%%%%%%%%%%%%%%%%%%%%%%%%%%%%%%%%%%%%%%%%%%%%%%%%%%%%%%%
\subsection{The disc and the cone}\label{sec:disco}
\begin{figure}
\subfloat[\label{fig:cono2d}]{\includegraphics[width=0.4\columnwidth,valign=c]{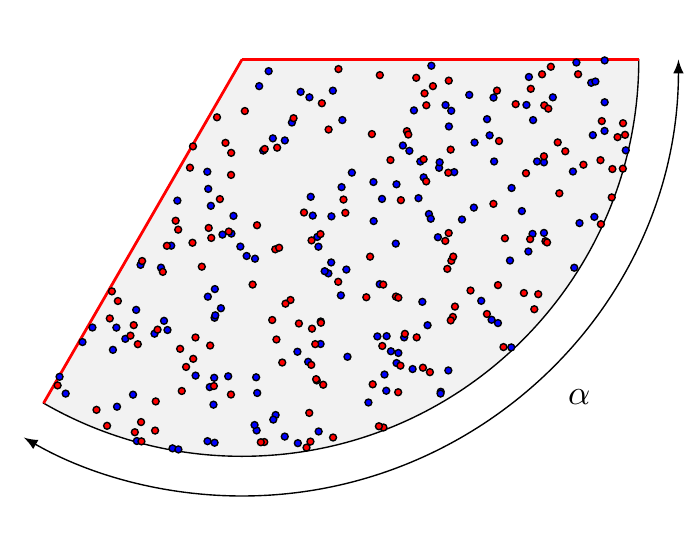}}\hfill
\subfloat[\label{fig:cono}]{\includegraphics[width=0.2\columnwidth,valign=c]{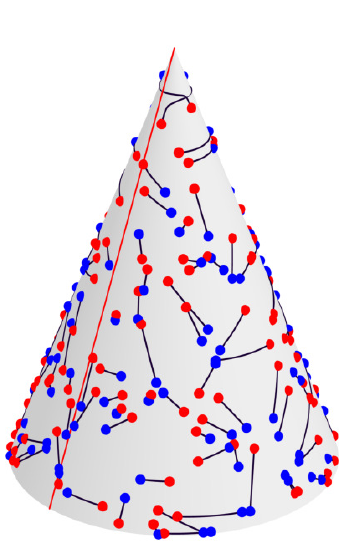}}\hfill
\subfloat[\label{fig:conodati}]{\includegraphics[width=0.4\columnwidth,valign=c]{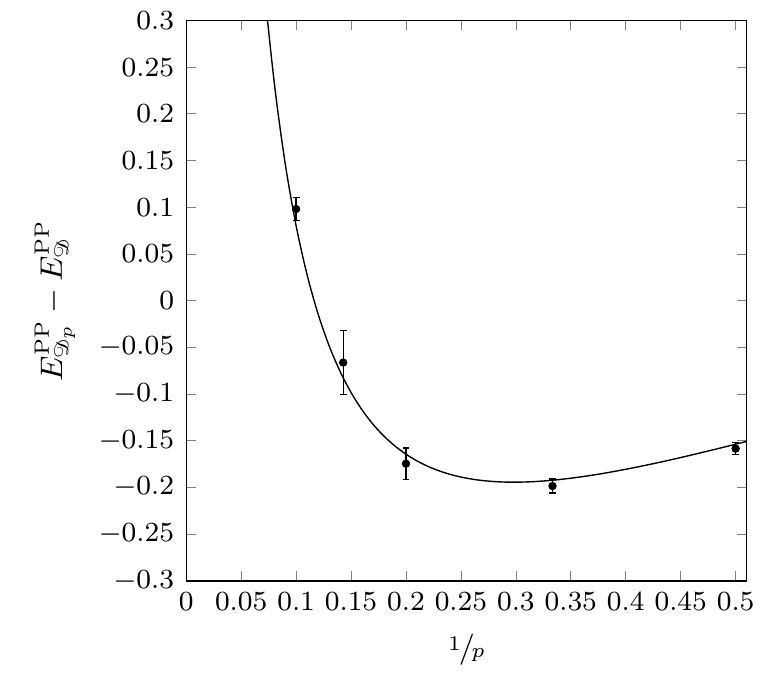}}
\caption{Random points on a circular sector (a) and optimal assignment on the corresponding cone (b) (in red, the part of the domain boundary where periodic boundary conditions are imposed). (c) Comparison between numerical results and our theoretical prediction for $E^{\rm PP}_{\mathcal D_p}-E^{\rm PP}_{\mathcal D}$.}
\end{figure}

Up to now, we have mostly solved the problem using the zeta regularization of the Laplacian, and relying on Kronecker's first limit formula. Only for the case of the torus, we have also performed the calculation of the Robin mass, and verified the prediction of the Morpurgo theorem.  In this section we will give the results for a geometry $\Omega$ in which the calculation of the Robin mass is done with relatively small effort, as the Green function can be guessed through the method of images, while the calculation of the Kronecker mass would require a sum over maxima and minima of Bessel functions.\footnote{This is because we have to consider Neumann boundary conditions for $\phi$. The sum would run on the zeros of Bessel functions if we had Dirichlet boundary conditions \cite{Elizalde1993}.} Let us introduce the notation $\mathcal D_p(r)$ for the circular sector of radius $r$ and angle $\frac{2\pi}{p}$, see Fig.~\ref{fig:cono2d},
\begin{equation}\label{settoredef}
\mathcal D_p(r)\coloneqq\left\{x\in\mathds C\colon|x|\leq r,\quad 0<\arg x<\frac{2\pi}{p}\right\}.
\end{equation}
The unit area condition implies $2\pi r^2=p$. We considered the case $p\in\mathds N$, and we choose periodic boundary conditions in the angular direction: this is equivalent to say that we identify the two radii of the sector, obtaining in this way a cone of height $r\sqrt{1-p^{-2}}$, see Fig.~\ref{fig:cono}. This surface is interesting, as it is the first example in our list of a surface with singular curvature, the conical singularity being at the vertex of the cone. We have argued in the introduction that, because of the scaling $\sim \sqrt{N^{-1}\ln N}$ of the field $\mu$, we expect that the same theory applies to the flat Euclidean space and to curved manifolds, \emph{as long as the curvature is non-singular}. The case of surfaces with a finite number of conical singularities would require a different (although feasible) argument, and the verification of our theory on this family of surfaces (as well as the surfaces treated in Section \ref{sec:sfera}) is an important validation of our predictions.

The Robin's mass for this case is obtained in Appendix \ref{app:sector}, and it is equal to
\begin{equation}
R_{\mathcal D_p}=-\frac {\ln\pi}{4\pi} 
+\frac {5p-2}{8\pi}
+\frac{\gamma_{\rm E}+\psi(\sfrac{1}{p})}{2\pi}-\frac{\ln p}{4\pi},
\end{equation}
where $\phi(z)$ is the digamma function. In particular, for $\alpha\to 2\pi$ we recover the case of the unit disc $\mathcal D\equiv \mathcal D_{1}$:
\begin{equation}
R_{\mathcal D}= \frac{1}{\pi} \left( \frac{3}{8} -\frac{\ln\pi}{4} \right).
\end{equation}
The Kronecker's mass is readily obtained using equation \eqref{diffRK}.

%%%%%%%%%%%%%%%%%%%%%%%%%%%%%%%%%%%%%%%%%%%%%%%%%%%%%%%
\subsection{The unit sphere and the real projective sphere}\label{sec:sfera}
An example of transportation problem on the surface of the sphere $\mathcal S^2$ has been considered in \cite{Peres18}, where the problem of transporting a uniform mass distribution into a set of random points on $\mathcal S^2$ is analyzed. As an example of applications of our approach to non-flat manifolds, here we consider the problem in our usual setting, i.e., a transportation between two atomic measures of random points. As in the previous cases, the information on the finite-size corrections is related to the spectrum of the Laplace--Beltrami operator on the manifold. It is well known that the eigenfunctions of $-\Delta$ on the surface of a sphere of radius $r$ are the spherical harmonics $Y_{l,m}(\theta,\phi)$ with $l\in \mathds{N}$ and $m\in \mathds{Z}$ with $-l\leq m\leq l$. The corresponding eigenvalues are
\begin{equation}
\lambda_{l,m}=\frac{l(l+1)}{r^2},
\end{equation}
that is, the eigenvalue $\frac{l(l+1)}{r^2}$ has multiplicity $2l+1$, for the range of integers $-l \leq m \leq l$.

We fix the surface area of the sphere to $1$ (taking $r=(4\pi)^{-\sfrac{1}{2}}$) and we proceed using the zeta regularization, computing
\begin{equation}
Z(s) = \frac{1}{(4 \pi)^s} \sum_{l \geq 1} \frac{2 l+1}{[ l (l+1)]^s}.
\end{equation}
In this case, after some algebra, we are led to use `just' the version of the zeta regularization for the Riemann zeta function (which is much simpler than the Kronecker formula)
\begin{equation}
\zeta(s)\coloneqq \sum_{k\geq 1} \frac{1}{k^s}=\frac{1}{s-1} + \gamma_{\text{E}} + \mathcal O(s-1)
\end{equation}
and we obtain
\begin{equation}
Z(s)=\frac{1}{4 \pi (s-1)}  - \frac{\ln(4 \pi)}{4\pi}  +\frac{\gamma_{\text{E}}}{2 \pi} - \frac{1}{4 \pi} + \mathcal O(s-1).
\end{equation}
The Kronecker mass for the unit sphere is therefore
\begin{equation}
\label{eq.KmassS2}
K_{\mathcal S^2} = -\frac{1+\ln(4 \pi) }{4\pi} +\frac{\gamma_{\text{E}}}{2 \pi}.
\end{equation}
Alternatively, we can use one of the regularizations illustrated in Section~\ref{ssec.kronetheo}. A convenient one is the evaluation of $W^{(\sfrac{1}{2})}$, which gives
\begin{equation}
\label{WfunRsphere}
\begin{split}
W^{(\sfrac{1}{2})}_{\mathcal S^2}(\epsilon)&=r^2\sum_{l \in \mathds{N}^+}\frac{2l+1}{l(l+1)} \e^{-\epsilon r^{-1}\sqrt{l(l+1)}}\\
& =r^2 \sum_{l\in\mathds N^+}\left(\frac{1}{l}+\frac{1}{l+1}\right)\e^{-\epsilon r^{-1}\sqrt{l(l+1)}}
\\&= r^2 \left(2\ln\frac{r}{\epsilon} - 1\right)+\mathcal O(\epsilon).\end{split}
\end{equation}
Recalling that in our case $r^2=(4\pi)^{-1}$, the final result is
\begin{equation}
\label{WfunRsphere2}
W^{(\sfrac{1}{2})}_{\mathcal S^2}(\epsilon) = -\frac{\ln\epsilon}{2\pi}-\frac{\ln(4\pi)+1}{4\pi}+ \mathcal{O}(\epsilon),
\end{equation}
that, in light of \eqref{diffWK}, allows to rederive equation \eqref{eq.KmassS2}.

\label{ssec.sfera}
\begin{figure}
\subfloat[\label{fig:lunea}]{\includegraphics[height=0.4\columnwidth,valign=c]{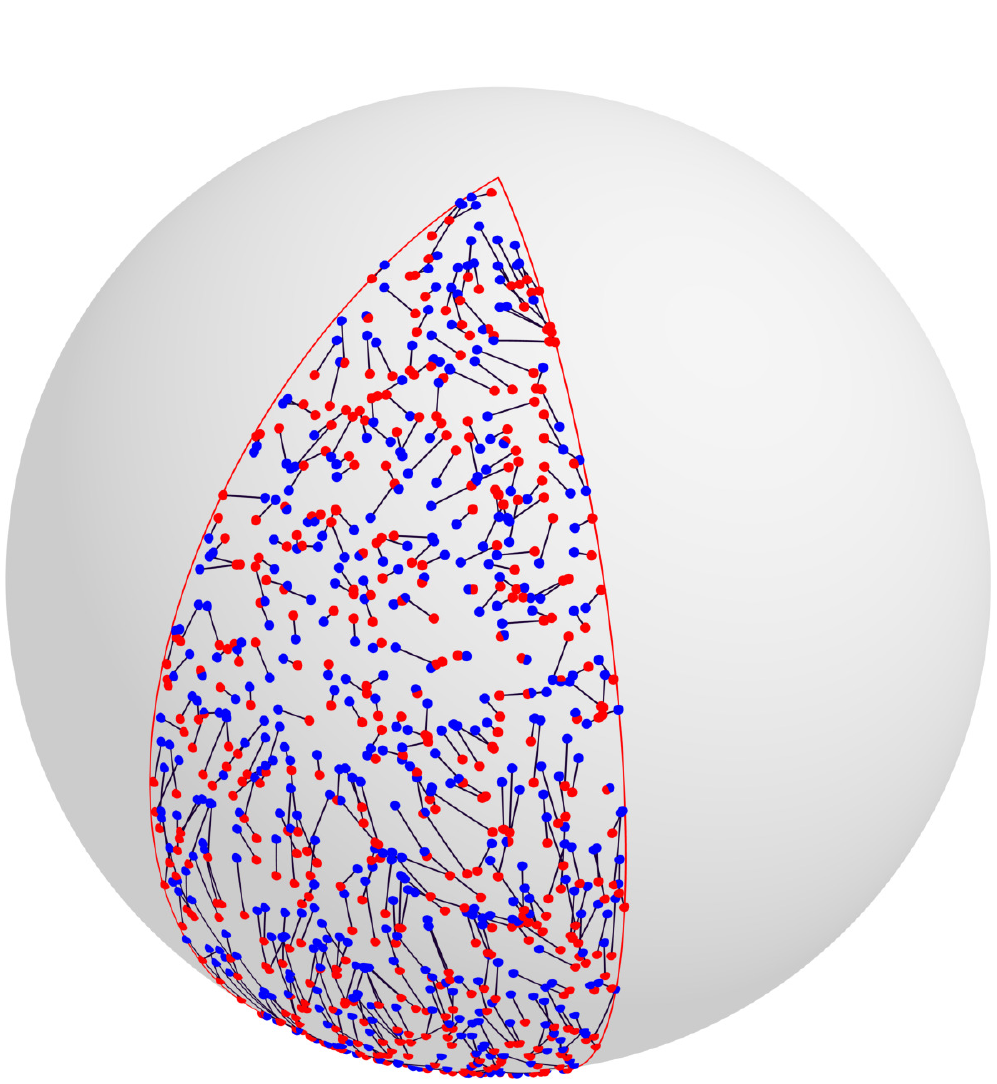}} 
\hspace{1cm}
\subfloat[\label{fig:spicchiodati}]{\includegraphics[height=0.4\columnwidth,valign=c]{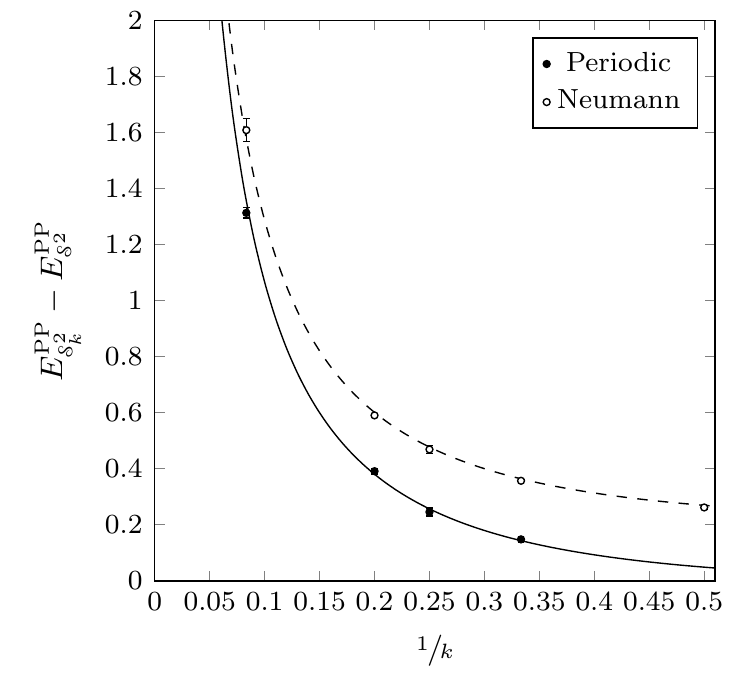}}
\caption{(a) Optimal assignment on a spherical lune. (b) Comparison between numerical results and our theoretical prediction for $E^{\rm PP}_{\mathcal S^2_k}-E^{\rm PP}_{\mathcal S^2}$ with periodic (smooth line) and Neumann (dashed line) boundary conditions. \label{fig:lune}}
\end{figure}

\paragraph{The spherical lune} The calculation above can be extended to the spherical lune $\mathcal S_k^2$, a surface on a sphere of radius $r$, $4\pi r^2=k$, contained by two half great circles which meet at antipodal points with dihedral angle $\frac{2\pi}{k}$, see Fig.~\ref{fig:lune}. In Appendix \ref{app:wedge} it is shown that the Kronecker's mass corresponding to this manifold is
\begin{equation}
K_{\mathcal S_k^2}=\frac{k-2-\ln(2\pi k)}{4\pi}+\frac{\gamma_{\rm E}}{2\pi}.
\end{equation}
Choosing periodic boundary conditions on the two half great circles, the Kronecker's mass takes an additional $-\frac{1}{2\pi}\ln 2$ contribution, and we obtain
\begin{equation}
K_{\mathcal S_k^2}=\frac{k-2-\ln(4\pi k)}{4\pi}+\frac{\gamma_{\rm E}}{2\pi},
\end{equation}
that reduces to \eqref{eq.KmassS2} for $k=1$, as it should.

\paragraph{The projective sphere}A variation of the problem on the (unit) sphere is the problem on the (unit) real projective sphere $\mathcal{PS}^2$, that is, the sphere in which antipodal points are identified. The eigenfunctions of the Laplace-Beltrami operator are still the spherical harmonics $Y_{l,m}(\theta,\phi)$ with $l\in \mathds{N}$ and $m\in \mathds{Z}$, $-l\leq m\leq l$, but we have to restrict ourselves to eigenfunctions that are invariant under the transformation $(\theta,\pi) \mapsto (\pi - \theta,\phi+ \pi)$, i.e., to even values of $l$. Working on the unit-area manifold accounts to have $2 \pi r^2=1$. We get
\begin{equation}
Z(s) =  \frac{1}{(2\pi)^s} \sum_{l \geq 1} \frac{4l+1}{[2l (2l+1)]^s}= \frac{1}{4 \pi (s-1)}  - \frac{\ln(2 \pi)}{4\pi}+\frac{\gamma_{\text{E}}}{2 \pi} - \frac{1}{2 \pi} + \mathcal O(s-1)
\end{equation}
so that the Kronecker's mass is
\begin{equation}
\label{eq.KmassSP2}
K_{\mathcal{PS}^2} = - \frac{\ln(2 \pi)}{4\pi}  + \frac{\gamma_E}{2 \pi} - \frac{1}{2 \pi}.
\end{equation}
Using a sharp cut-off instead
\begin{equation}
\label{WfunRspherep}
\begin{split}
W^{\rm sharp}_{\mathcal{PS}^2}(\epsilon)&=r^2\sum_{l \in \mathds{N}^+}\frac{4l+1}{2l(2l+1)}\theta\left(\frac{1}{\epsilon}-\frac{2l(2l+1)}{r^2}\right)\\
&=r^2 \sum_{l\in\mathds N^+}\left(\frac{1}{2l}+\frac{1}{2l+1}\right)\theta\left(\frac{1}{2} \sqrt{\frac{r^2}{\epsilon}+\frac{1}{4}}-\frac{1}{4}-l\right)\\
&=r^2 \left[H\left(\frac{r}{\sqrt{\epsilon}}\right) -1\right]+\mathcal{O}(\epsilon)=\frac{1}{2 \pi}\left(-\frac{1}{2}\ln \epsilon-\frac{\ln(2 \pi)}{2}+\gamma_{\rm E}-1 +\mathcal{O}(\epsilon)\right)
\end{split}
\end{equation}
which, again by using \eqref{WfunlimR} and \eqref{diffWKR}, allows to rederive equation \eqref{eq.KmassSP2}.

%%%%%%%%%%%%%%%%%%%%%%%%%%%%%%%%%%%%%%%%%%%%%%%%%%%%%%
\section{Numerical results}\label{sec:numerica}
We have numerically investigated all the cases described above to check our predictions. To solve the assignment problem we have used an implementation of the Jonker-Volgenant algorithm~\cite{Jonker1987}. For each domain $\Omega$, we have computed the expected optimal cost averaging over at least $10^4$ independent instances and different sizes $N$ of the system, $32\leq N\leq 1024$. In each case, we have estimated $c^{\rm \bullet P}_\Omega(N)$ assuming that they are indeed constant at the leading order, i.e., $c^{\rm \bullet P}_\Omega(N)\equiv c^{\rm \bullet P}_\Omega$, via a least square regression. Here $\bullet=\{\rm P,S,T,F,U\}$. Our results are given in Table~\ref{datiPP}. 

For each domain we have also computed $c_*^{\rm \bullet P}=c_\Omega^{\rm \bullet P}-K_\Omega$ that we expect to be domain-independent. In the PP case, our best estimation of $c_*^{\rm PP}$, obtained for the PP problem on the torus (see Fig.~\ref{fig:ct}) is
\begin{equation}
c_*^{\rm PP}=c_{\Omega}^{\rm PP}-K_\Omega = 0.29258(2).
\end{equation}
Numerically, however, we cannot rule out a weak $N$-dependence in $c_*^{\rm PP}$. In a similar way we have obtained the results given in \eqref{cstarxp}, that we repeat here,
\begin{align}
c_*^{\rm SP}&=0.4156(5)
&
c_*^{\rm TP}&=0.413(2)
&
c_*^{\rm UP}&=0.4038(3).
\end{align}

To verify our results in Eqs.~\eqref{costoppxp} we have also proceeded in this way. Given two unit-area domains $\Omega$, $\bar\Omega$, and a given type of the problem (PP, SP, etc), we have computed, for each $N$, $\Delta E^{\rm \bullet P}_{\Omega,\bar\Omega}(N)=c^{\rm \bullet P}_{\Omega}(N)-c^{\rm \bullet P}_{\bar\Omega}(N)$ and then extrapolated to $N\to +\infty$. If the arguments above are correct, then
\begin{align}
\Delta E^{\rm PP}[\Omega,\bar\Omega]&\coloneqq \lim_{N\to +\infty}\Delta E^{\rm PP}_{\Omega,\bar\Omega}(N)=2K_\Omega-2K_{\bar \Omega},\\
\Delta E^{\rm \bullet P}[\Omega,\bar\Omega]&\coloneqq \lim_{N\to +\infty}\Delta E^{\rm \bullet P}_{\Omega,\bar\Omega}(N)=K_\Omega-K_{\bar \Omega},\quad \bullet=\{\rm S,T,F,U\}.
\end{align}
In Fig.~\ref{fig:qtb} we plot in particular 
\begin{align}
\Delta E^{\rm\bullet P}_{\mathcal R}(\rho)&\coloneqq \Delta E^{\rm \bullet P}[\mathcal R(\rho),\mathcal R(1)],\\
\Delta E^{\rm\bullet P}_{\mathcal T}(\rho)&\coloneqq \Delta E^{\rm \bullet P}[\mathcal T(i\rho),\mathcal T(i)],\\
\Delta E^{\rm\bullet P}_{\mathcal B}(\rho)&\coloneqq \Delta E^{\rm \bullet P}[\mathcal B(\rho),\mathcal B(1)],
\end{align}
as functions of $\rho$ for $\bullet=\{\rm P,S\}$. Similarly, in Fig.~\ref{fig:cmk} we present our results for (the absolute value of)
\begin{align}
\Delta E^{\rm\bullet P}_{\mathcal C}(\rho)&\coloneqq \Delta E^{\rm \bullet P}[\mathcal C(\rho),\mathcal C(1)],\\
\Delta E^{\rm\bullet P}_{\mathcal M}(\rho)&\coloneqq \Delta E^{\rm \bullet P}[\mathcal M(\rho),\mathcal M(1)],\\
\Delta E^{\rm\bullet P}_{\mathcal K}(\rho)&\coloneqq \Delta E^{\rm \bullet P}[\mathcal K(\rho),\mathcal K(1)]
\end{align}
for $\bullet=\{\rm P,S\}$.\footnote{Of course, all the signs just come out as predicted.} In Fig.~\ref{fig:conodati} and in Fig.~\ref{fig:spicchiodati} we have also considered the differences of average optimal costs in the case of the circular sector and of the spherical lune. In all investigated cases we found a perfect agreement with our predictions. 

\begin{table}
\resizebox{0.8\columnwidth}{!}{
\begin{tabular}{p{0.1\columnwidth}l|ll|lll}
\hline\hline
 & \qquad $K_\Omega$ & \quad $c^{\rm PP}_{\Omega}$ & $c^{\rm PP}_{\Omega}-K_\Omega$& \quad $c^{\rm \bullet P}_{\Omega}$ & $c^{\rm \bullet P}_{\Omega}-K_\Omega$&Grid\rule{0pt}{11pt}\\
 \hline
$\mathcal T$ &$-0.2270289\dots$& $0.06555(2)$&$0.29258(2)$& $0.1883(3)$&$0.4154(3)$&Square\\
$\mathcal T(\e^{\sfrac{\pi i}{3}})$ &$-0.2287134\dots$ & $0.064(2)$&$0.293(2)$ & $0.184(2)$&$0.413(2)$&Triangle \\
$\mathcal R$ &$\hphantom{-}0.0499556\dots$ & $0.3420(3)$&$0.2921(3)$& $0.460(4)$&$0.410(4)$&Square\\
$\mathcal C$ &$-0.1026239\dots$& $0.1895(3)$&$0.2921(3)$& $0.310(2)$&$0.412(2)$&Square\\
$\mathcal M$ &$-0.1302033\dots$& $0.1626(3)$&$0.2928(3)$& $0.284(3)$&$0.414(3)$&Square\\
$\mathcal K$ &$-0.2276239\dots$& $0.0646(8)$&$0.2922(8)$& $0.1880(5)$&$0.4156(5)$&Square\\
$\mathcal B$ &$-0.2000444\dots$& $0.0925(1)$&$0.2926(2)$& $0.216(1)$&$0.416(1)$&Square\\
$\mathcal D$ &$\hphantom{-}0.0098204\dots$& $0.302(1)$&$0.292(1)$& $0.423(3)$&$0.413(3)$&Fibonacci\\
$\mathcal S^2$ &$-0.1891233\dots$& $0.1034(2)$ &$0.2925(2)$& $0.2255(8)$ &$0.4146(8)$&Fibonacci\\
$\mathcal{PS}^2$ &$-0.2135418\dots$& $0.079(1)$ & $0.292(1)$& $0.2022(8)$ & $0.4157(8)$&Fibonacci\\
\hline\hline
\end{tabular}}\vspace{0.5cm}
\caption{Kronecker mass and finite-size corrections $c^{\rm PP}_\Omega$ evaluated by  numerical simulations of random assignments on different domains. An estimation of $c^{\rm PP}_\Omega-K_\Omega$ is also given. We also give our numerical results for $c_\Omega^{\rm \bullet P}$, obtained performing random assignments on the same domains but fixing one set of points on a grid. The type of adopted grid is specified in the last column. \label{datiPP}}
\end{table}

\section{Conclusions and perspectives}\label{sec:conclusion}
In this paper we have considered the assignment problem between two sets of random points on a generic two-dimensional smooth manifold of unit area. We have showed, by means of analytical arguments and numerical simulations, that the average optimal cost can be written as
\begin{subequations}
\begin{equation}
E_{\Omega}(N)\coloneqq \frac{1}{2\pi}\ln N+2c_*^{\rm PP}(N)+2K_\Omega+o(1),
\end{equation}
if both sets of points are random (Poisson--Poisson case), and as
\begin{equation}
E_{\Omega}(N)\coloneqq \frac{1}{4\pi}\ln N+c_*^{\rm \bullet P}(N)+K_\Omega+o(1),\quad\bullet=\{\rm S,T,F,U\},
\end{equation}
\end{subequations}
if one of the two sets is fixed on a grid (square, triangular, Fibonacci) or replaced with the uniform measure. In the equations above, $K_\Omega$ is a precise quantity that can be obtained by a zeta-regularization of the trace of the inverse Laplace--Beltrami operator on $\Omega$. The contributions $c^{\rm\bullet P}_*$ are instead independent on $\Omega$ and related to the `local details' of the problem (i.e., if the assignment is between random points, or with a grid, or with the uniform measure). We have given an exact computation of $K_\Omega$ for different domains, and using different procedures. 

The quantity $c_*^{\rm \bullet P}(N)$ shows a weak dependence on $N$ (if no dependence at all): it has been proven indeed that $c_*^{\rm UP}=\mathcal{O}(\sqrt{\ln N\ln\ln N})$ \cite{Ambrosio2018}, a bound that, because of triangular inequality, holds for all the cases that we have considered. Assuming that $c_*^{\rm \bullet P}(N)$ are constants, we have verified, within our numerical precision, their independence on $\Omega$ in all considered transportation cases (Poisson--Poisson, grid--Poisson, uniform--Poisson).

Our results reduce the computation of the (leading) finite-size correction to the optimal cost to the computation of the $\Omega$-independent contributions $c_*^{\rm \bullet P}(N)$. These contributions are intrinsically dependent on the local nature of the problem (and therefore change if, for example, we fix on a grid one of the two sets of points) and are inherited by the regularization of the highest part of the spectrum of $-\Delta$, as discussed in Section~\ref{sec:teoria}. What are the properties (and possibly the exact value, if they are constant) of $c_*^{\rm \bullet P}(N)$ remains an open question.

Finally, analogous results are expected to hold in higher dimension at the leading order. In particular, the analysis in \cite{Caracciolo:158} suggests that, for $d > 2$, the local properties of the problem affect the coefficient of the leading term, with a finite-size correction depending on the spectrum of the Laplacian only.
This case may also be investigable with our tools, as indeed versions of the Weyl law, which we crucially use in Section \ref{sec:regola}, exist in generic dimension. We leave the investigation of the higher dimensional problem for future works.

\subsection*{Acknowledgement}
E.~Caglioti and G.~Sicuro would like to thank Giorgio Parisi for putting them in contact. D.~Benedetto and E.~Caglioti thanks Gabriele Mondello and Riccardo Salvati Manni for clarifying discussions about the case of the torus. The authors are grateful to J\"{u}rg Fr\"{o}hlich for his careful reading of the manuscript. A.~Sportiello is partially supported by the Agence Nationale de la Recherche, Grant Number ANR-18-CE40-0033 (ANR DIMERS).

\appendix

\section{Comparison of GP problem with UP problem on the flat torus}\label{app:GPUP}
As commented in the main text, the very same arguments presented for the PP case in Section \ref{sec:teoria} can be repeated for the GP case and the UP case, the only difference being that \eqref{eq.noise} has to be replaced by
\begin{equation}
\mathbb E[\delta\nu (x)\delta\nu (y)]=\frac{1}{N}\left(\delta_y(x)-1\right),
\end{equation}
The result is an overall $\frac{1}{2}$ factor, as shown in the final formula \eqref{costoxp}. There is however no guarantee that $c_\Omega^{\rm PP}(N)=c_\Omega^{\rm \bullet P}(N)$ for $\bullet=\{\rm S,T,F,U\}$ at fixed $\Omega$ as one might naively expect from \eqref{costoppxp}, because these quantities depends on the regularizing function $F^{\rm \bullet P}(\lambda/N)$, that, even assuming that it exists, is expected to be in general different in each case. In \cite{Ambrosio2016} it is proved that
\begin{equation}\label{inambrosio}
c_\Omega^{\rm PP}(N)\leq c_\Omega^{\rm UP}(N).
\end{equation}
This equation, combined with \eqref{costoppxp}, implies
\begin{equation}
c_*^{\rm PP}(N)\leq c_*^{\rm UP}(N).
\end{equation}
As discussed in Section \ref{sec:teoria}, one way to numerically estimate $c_*^{\rm UP}(N)$ is to perform a transportation between two sets of points, supposing that one of them (e.g., the red ones) is fixed on a grid and not random. As intuitively expected, a grid approximation provides some information on $c_*^{\rm UP}$. 

For example, let us consider the transportation between a set $\mY=\{Y_i\}_{i=1,\dots,N}$ of random points on the (flat) torus $\mathcal T$ and a set of $hN=L^2$ points $\mX=\{X_i\}_{i=1,\dots,hN}$ fixed on a square grid. This is \textit{not} an assignment problem because the cardinality of the two sets is different. However, if $h\in\mathds N$ the transport can be obtained as an assignment ``replicating'' each point in $\mY$ $h$ times, see Fig.~\ref{fig:torogrid}, so that in the optimal configuration each point of the original set $\mY$ will be associated to $h$ grid points. By classical convexity properties of the squared Wasserstein distance, it can be proved that, given $hN=L^2$ and considering a squared $L\times L$ grid on the unit flat $2$-torus,
\begin{equation}\label{UG}
c^{\rm GP}_{\mathcal T}(N)-\frac{1}{6h}\leq c^{\rm UP}_{\mathcal T}(N)\leq c_{\mathcal T}^{\rm GP}(N),
\end{equation}
where we denoted by a generic GP the correction in the discrete transportation problem (in particular, $c^{\rm GP}_{\mathcal T}(N)=c^{\rm SP}_{\mathcal T}(N)$ for $h=1$). By means of the GP problem, $c^{\rm UP}_{*}(N)$ can be estimated for each $N$ in the limit $h\to +\infty$.

To prove Eq.~\eqref{UG}, first we notice that if a probability measure $\nu_1$ is the convex combination of measures $\nu_{\eta}$,  that is, if we can write
\begin{equation}
\nu_1=\int\nu_{\eta}\dd\gamma(\eta)
\end{equation}
where $\gamma(\eta)$ is a probability measure, then, for any measure $\nu_2$
\begin{equation}
W_2^2(\nu_1,\nu_2)\leq \int W_2^2(\nu_\eta,\nu_2)\dd\gamma(\eta) . 
\end{equation}
Let us assume now that the following empirical measure is given on the torus $\mathcal T$,
\begin{equation}
\nu_{\mX}=\frac{1}{hN}\sum_{i=1}^{hN}\delta_{X_i}
\end{equation}
concentrated on $hN=L^2$ points on the torus, $L\in\mathds N$, and let us assume that such points $\mX\coloneqq \{X_i\}_{i=1,\dots,hN}$ lie on a square grid of step $L^{-1}$. Let us denote also with $\mY=\{Y_i\}_{i=1,\dots N}$ a set of $N$ points sampled from the uniform distribution, and  with $\nu_{\mY}$ their empirical density. Noticing that the uniform measure on the torus is a convex combination of grid measures in the torus, we get, for any $N$,
\begin{equation}
E^{\rm UP}_{\mathcal T}(N)\leq E^{\rm GP}_{\mathcal T}(N),
\end{equation}
where we label by GP the quantities corresponding to the grid-Poisson transportation problem. Now let us estimate $E^{\rm UP}_{\mathcal T}(N)$ from below. In the optimal solution for $W_2^2(\sigma,\nu_{\mY})$ the point $x$ is joined, for almost all $x$, with one of the points in $\mY$. Let us denote with $Y_{J(x)}\in\mY$ the point in $\mY$ to which $x$ is associated. Moreover we can associate almost every point $x$ to the closest grid point, let us denote it by $X_{I(x)}\in\mX$. Using the functions $I(x)$ and $J(x)$ we can build a coupling between $\nu_{\mY}$ and $\nu_{\mX}$ defined by 
\begin{equation}
\label{GPapprox}
P_{i,j}\coloneqq \int \dd x\ \delta_{i,I(x)} \delta_{j,J(x)}.
\end{equation}
In fact, it is easy to check that $\forall i$ $\sum_u P_{i,u}=N^{-1}$ and, $\forall j$, $\sum_u P_{u,j}=(hN)^{-1}$. Therefore, being $W_2^2(\nu_{\mX},\nu_{\mY})$ the minimum on all the possible coupling between $\nu_{\mX}$ and $\nu_{\mY}$ we get
\begin{multline}
W_2^2(\nu_{\mX},\nu_{\mY})\leq\sum_{i,j}P_{i,j} |X_i-Y_j|^2
=\int \dd x\ |X_{I(x)}-Y_{J(x)}|^2\\
=\int \dd x\left[|X_{I(x)}-x|^2+|x-Y_J(x)|^2\right]-\int \dd x\left[2(x-X_{I(x)})(x-Y_{J(x)})\right].
\end{multline}
Now, taking the average on the locations of the $\mY$ points, and noticing that $\mathbb E[x-Y_{J(x)}]=0$,
we get
\begin{equation}
\label{intermedia1}
\mathbb E\left[ W_2^2(\nu_{\mX},\nu_{\mY})\right]\leq \int\dd x\ |X_{I(x)}-x|^2 + W_2^2(\sigma,\nu_{\mY}).
\end{equation}
The points $x$ joined with $X_i$ by means of $I(x)$ are the points in a square of side $(hN)^{-\sfrac{1}{2}}$ centered in $X_i$, therefore
\begin{equation}
\int\dd x\ |X_{I(x)}-x|^2=hN\int_{\mathclap{\left[0,\sfrac{1}{\sqrt{hN}}\right]^2}}|x|^2\dd x=\frac{1}{6hN},
\end{equation}
so that Eq.~\eqref{intermedia1} becomes
\begin{equation}
\mathbb E\left[ W_2^2(\nu_{\mX},\nu_{\mY})\right]\leq \frac{1}{6hN}+ W_2^2(\sigma,\nu_{\mY}),
\end{equation}
i.e., $E^{\rm UP}_{\mathcal T}(N)\geq E^{\rm GP}_{\mathcal T}(N)-\frac{1}{6h}$ and therefore
\begin{equation}
c_{\mathcal T}^{\rm UP}\geq c_{\mathcal T}^{\rm GP}-\frac{1}{6h}.
\end{equation}
As expected, $\lim_{h\to +\infty}c_{\mathcal T}^{\rm GP}=c_{\mathcal T}^{\rm UP}$.

%%%%%%%%%%%%%%%%%%%%%%%%%%%%%%%%%%%%%%%%%%%%%%%%%%%%%%%%%%%%%%%%%%%%%%%%%%%%%%
\section{Robin's mass for the circular sector}\label{app:sector}
In this appendix we compute the Robin mass of the Laplace--Beltrami Green function on the circular sector $\mathcal D_p$ of angle $\alpha=\frac{2\pi}{p}$, with $p\in\mathds N$, with periodic boundary conditions in the angular direction. We will work on the sector $\mathcal D_p(r)$ defined in Eq.~\eqref{settoredef} and we will then restrict ourselves to the unit area case. Let us start considering the functions
\begin{subequations}
\begin{align}
L(x) &\coloneqq -\frac 1{2\pi} \ln |x|,\\
L^e(x,y) &\coloneqq -\frac 1{2\pi} \ln \left|x -\frac{ y }{|y|^2}r^2\right|,\\
A^{(p)}(x,y)&\coloneqq\sum_{k=0}^{p-1}L(x- R_{k\alpha}y)+\sum_{k=0}^{p-1}L^e(x, R_{k\alpha}y),
\end{align} 
\end{subequations}
where $R_{\theta}$ is the rotation matrix of an angle $\theta$ around the origin. The function $A^{(p)}(x,y)$ is such that $A^{(p)}(R_{\alpha}x,y)=A^{(p)}(x,y)$. In the circular sector $\mathcal D_p(r)$ we have
\begin{subequations}
\begin{align}
-\Delta_y A^{(p)}(x,y)=\delta(x - y),\\
\partial_nA^{(p)}(x,y)=-\frac{p}{2\pi r},
\end{align} 
\end{subequations}
where $\left.\partial_{n}A^{(p)}(x,y)\right|_{|y|=r}$ is the normal derivative in $x$ with respect to the boundary $|y|=r$ of the domain. The function
\begin{equation}
g(x,y) = A^{(p)}(x, y) + \frac {p}{4\pi r^2} |x|^2 
\end{equation}
is therefore the Green function of the Laplacian on $\mathcal D_p$ with Neumann boundary conditions on $|y|=r$. To impose \eqref{zeroarea} we compute
\begin{equation}
c(y) =\frac{p}{\pi r^2} \int_{\mathcal D_p(r)} g(z,y) \dd z= \frac{1}{\pi r^2} \int_{\mathcal D(r)} g(z,y) \dd z.
\end{equation}
Observe that  $g(x,y)$ is periodic in $y$ and therefore $c(y)$ is periodic as well. The Green function is therefore
\begin{equation}
G(x, y) = g(x,y) -c(y).
\end{equation}
For $y\to x$ we can write
\begin{equation}
G(x,y)=-\frac{1}{2\pi}\ln|x-y|+\gamma(x,y), 
\end{equation}
with regular part $\gamma(x,y)$ given by
\begin{equation}\label{eq.app.gamma}
\gamma(x, y) \coloneqq\sum_{k=1}^{p-1} L(x - R_{k\alpha}y)+\sum_{k=0}^{p-1} L^e(x,R_{k\alpha}y) + \frac p{4\pi r^2} |x|^2
-c(y).\end{equation}
To compute the Robin mass we have to estimate
\begin{equation}
\int_{\mathcal D_p(r)} \gamma(x,x)\dd x=\frac{1}{p}\int_{\mathcal D(r)} \gamma(x,x)\dd x=I_1+I_2+\frac{1}{p}\sum_{k=0}^{p-1}I_3(k)+\frac{1}{p}\sum_{k=1}^{p-1}I_4(k).
\end{equation}
where the four types of contributions that appear in the equation above, associated to the summands on the RHS of \eqref{eq.app.gamma}, are defined as
\begin{subequations}
\begin{align}
I_1&=-\frac 1{\pi r^2} \iint_{\mathclap{\mathcal D(r)\times \mathcal D(r)}} L^e(z,x)\dd z\dd x,\\
I_2&=-\frac 1{\pi r^2} \iint_{\mathclap{\mathcal D(r)\times \mathcal D(r)}}L(z -x)\dd z\dd x,\\
I_3(k)&=\int_{\mathcal D(r)}  L^e(x,R_{k\alpha}x)\dd x,\quad \text{with }k=0,\dots p-1,\\
I_4(k)&=\int_{\mathcal D(r)} L(x - R_{k\alpha}x)\dd x,\quad \text{with }k=1,\dots,p-1.
\end{align}
\end{subequations}
We will use the identities
\begin{equation}
\int_0^r x \ln x \dd x = \frac {r^2}{2} \ln r - \frac {r^2}4, \qquad \int_0^1 x^3 \ln x  \dd x = -\frac{1}{16}. 
\end{equation}
Let us start from
\begin{equation}
I_1=-\frac 1{\pi r^2} \iint_{\mathclap{\mathcal D(r)\times \mathcal D(r)}} L^e(z,x)\dd x\dd z=-\frac 1{\pi r^2} \iint_{\mathclap{\mathcal D(r)\times \mathcal D(r)}} \ln\left|z-\frac{x}{|x|^2}r^2\right|\dd x\dd z.
\end{equation}
Here the key observation is that $|x|^{-1}r^2>r$, i.e., $x|x|^{-2}r^2$ always lies outside $\mathcal D(r)$, and therefore the integral in $x$ is equal to the value of the integrand for $x=0$ times the area of $\mathcal D(r)$,
\begin{equation}
I_1=-\int_{\mathcal D(r)} L^e(0,x)\dd x
= \int_0^r x\ln \frac{r^2}{x}\dd x = \frac {r^2}2\ln r + \frac {r^2}4.
\end{equation}

Similar arguments help us to evaluate $I_2$,
\begin{equation}
\begin{split}
 I_2&=-\frac 1{\pi r^2} \iint_{\mathclap{\mathcal D(r)\times \mathcal D(r)}}L(z-x)\dd z\dd x
 =-\frac 1{2\pi^2 r^2} \int_{\mathclap{\mathcal D(r)}}\dd x\Bigg[\quad\int_{\mathclap{\substack{z\in\mathcal D(r)\\|z|<|x|}}} \ln |z -x| \dd z+\int_{\mathclap{\substack{z\in\mathcal D(r)\\|z|>|x|}}} \ln |z -x| \dd z\Bigg]\\
 &=-\frac 1{2\pi^2 r^2} \int_{\mathclap{\mathcal D(r)}}\dd x\Bigg[\pi |x|^2 \ln |x|+\int_{\mathclap{\substack{z\in\mathcal D(r)\\|z|>|x|}}} \ln |z| \dd z\Bigg]=-\frac 1{\pi r^2} \int_{\mathclap{\mathcal D(r)}}|x|^2 \ln |x| \dd x=\frac{2}{r^2}\int_0^r x^3\ln x\dd x\\
 &=\frac{r^2\ln r}{2}-\frac{r^2}{8}.
\end{split} 
\end{equation}

The integrals $I_3(k)$ and $I_4(k)$ can be computed introducing a complex representation of the integration variable, $x= u\e^{i\vartheta}$, and then writing
\begin{subequations}
\begin{align}
|x - R_{k\alpha}x|^2&= 4u^2\sin^2\left(\frac{k\alpha}{2}\right),\\
\left|x - \frac{r^2}{|x|^2}R_{k\alpha}x\right|^2&=\frac{u^4 + r^4 - 2\cos\left(k\alpha\right) u^2 r^2}{u^2}.
\end{align}
\end{subequations}
After this change of variables, it is found that
\begin{equation}
I_4(k)= -\frac {r^2}2 \ln\left(2\sin(\pi k)\right)-\frac {r^2}2\ln r + \frac {r^2}4.
\end{equation}
Let us finally compute $I_3(k)$. Denoting by $a=\cos(k\alpha)$,
\begin{equation}
\begin{split}
I_3(k)&= -\frac 12 \int_0^r u \ln (u^4 + r^4 -2 au^2 r^2)\dd u + \int_0^r u \ln u \dd u\\
&= -\frac {2r^2\ln r+r^2}4 -\frac{r^2}{2} \int_0^1 u \ln ( 1+u^4 -2au^2)\dd u\\
&= -\frac {2r^2\ln r+r^2}4 -\frac{r^2}{4} \int_0^1 \ln ( 1+u^2 -2a u)\dd u\\
&=-\frac {2r^2\ln r+r^2}4 -\frac{r^2}{4}\ln(2-2a)+\frac{r^2}{2} \int_0^1\frac {u(u-a)}{u^2 - 2au + 1}\dd u.
\end{split}
\end{equation}
Using now the fact that
\begin{multline}
 \int_0^1 \frac {u(u-a)}{u^2 - 2ua + 1}\dd u=\int_0^1 \left[1 +
\frac{a}{2}\frac{\partial}{\partial u}\left[
\ln (u^2 -2 ua + 1)\right] +\frac {a^2 -1}{u^2 - 2ua +1}\right]\dd u\\
=1+\frac{a}{2}\ln(2-2a)-\frac{\pi-k\alpha}{2}\sin(k\alpha),
\end{multline}
we obtain
\begin{equation}
I_3(k) =-\frac {2r^2\ln r-r^2}4-\frac {r^2}{2} (1-\cos (k\alpha)) \ln\left(2\sin\left(\frac{k\alpha}{2}\right)\right)-r^2 \frac{\pi - k\alpha}{4} \sin(k\alpha).
\end{equation}
Summing all contributions, we obtain
\begin{multline}\label{eq.app.sett1}
\int_{\mathcal D_p(r)}\gamma(x,x)\dd x=\\
=\frac {r^2}{2p} \ln r+\frac {r^2}{8} \left( 5 - \frac 2p\right)+\frac {\alpha r^2}{4p}\sum_{k=1}^{p-1}k\sin(k\alpha)- \frac {r^2}{2p}\sum_{k=1}^{p-1} (2-\cos (k\alpha))\ln\left(2\sin\left(\frac{k\alpha}{2}\right)\right).
\end{multline}
The last sum can be simplified as follows:
\begin{equation}
\sum_{k=1}^{p-1} \left(2-\cos \left(k\alpha\right)\right)\ln\left(2\sin\left(\frac{\pi k}{p}\right)\right)=\ln(2p^2)-\sum_{k=1}^{p-1}\cos\left(k\alpha\right)\ln\sin\left(\frac{2\pi}{p}\right).
\end{equation}
Applying the Gauss's digamma theorem \cite{EMOT1981}
\begin{multline}
\sum_{k=1}^{p-1}\cos\left(k\alpha\right)\ln\sin\left(\frac{\pi k}{p}\right)=\\
=2\sum_{k=1}^{\mathclap{\lceil \sfrac{p}{2}\rceil-1}}\cos\left(\frac{2\pi k}{p}\right)\ln\sin\left(\frac{\pi k}{p}\right)=
\psi\left(\sfrac{1}{p}\right)+\gamma_{\rm E}+\ln(2p)+\frac{\pi}{2}\cot\left(\frac{\pi}{p}\right)
\end{multline}
we can obtain the final expression
\begin{equation}\label{eq.app.sett1}
\int_{\mathcal D_p(r)}\gamma(x,x)\dd x=\frac {r^2}{2p} \ln r
+\frac {r^2}{8} \left( 5 - \frac 2p\right)
+r^2\frac{\gamma_{\rm E}+\psi(\sfrac{1}{p})-\ln p}{2p}.
\end{equation}

To restrict ourselves to the case of unit area, we impose $\pi r^2 = p$ obtaining the searched Robin's mass
\begin{equation}
  \label{eq:fine}
 R_{\mathcal D_p}=-\frac {\ln\pi}{4\pi} 
+\frac {5p-2}{8\pi}
+\frac{\gamma_{\rm E}+\psi(\sfrac{1}{p})}{2\pi}-\frac{\ln p}{4\pi}.
\end{equation}
For $\alpha=2\pi$, i.e., $p=1$, we obtain the Robin's mass for the disc, 
\begin{equation}
 R_{\mathcal D}=\frac {3}{8\pi}-\frac {\ln\pi}{4\pi}.
\end{equation}

%%%%%%%%%%%%%%%%%%%%%%%%%%%%%%%%%%%%%%%%%%%%%%%%%%%%%%%%%%%%%%%%%%%%%%%%%%%%%%
\section{Kronecker's mass for the spherical lune}\label{app:wedge}
We consider the surface of the sphere but we wish to take only a portion $\mathcal S^2_k$ around the $z$ axis. Let us first observe that the eigenvectors of the Laplace--Beltrami operator on the sphere are the spherical harmonics,
\begin{equation}
Y_l^m(\theta,\phi)\propto \e^{im\phi} P_l^m(\cos\theta),\quad \phi\in[0,2\pi),\ \theta\in[0,\pi],\ l,m\in\mathds N_0\text{ with }-l\leq m\leq l.
\end{equation}
The eigenvector $Y^m_l(\theta,\phi)$ has eigenvalue $\frac{1}{r^2}l(l+1)$. Here $\theta$ is the colatitude and $\phi$ the longitude on the sphere, whereas $P_l^m(x)$ is an associated Legendre polynomial (we have omitted a normalization constant).

Let us now consider the lune $\mathcal S^2_k$ with periodic boundary conditions. This means that we restrict ourselves to the span of eigenvectors having values $m$ such that $m\equiv 0\mod k$. That is, the degeneracy in $m$ of the eigenvalue $l(l+1)r^{-2}$ of the Laplace--Beltrami operator, when $l=nk+r$ with $r=0,\dots,k-1$, is $2n+1$. The condition of unit area means $4\pi r^2 = k$, so that
\begin{equation}
Z(s)=\left(\frac{k}{4\pi}\right)^s\left[\sum_{r=1}^{k-1}\frac{1}{r^s(r+1)^s}+\sum_{r=0}^{k-1}\sum_{n=1}^\infty\frac{2n+1}{(nk+r)^s(nk+r+1)^s}\right].
\end{equation}
The contribution obtained for $n=0$ can be immediately summed,
\begin{equation}
 \left(\frac{k}{4\pi}\right)^s\sum_{r=0}^{k-1}\frac{1}{r^s(r+1)^s}=\frac{k-1}{4\pi}+\mathcal O(s-1).
\end{equation}
The singular part for $s=1$ comes from
\begin{equation}
2\left(\frac{k}{4\pi}\right)^s\sum_{r=0}^{k-1}\sum_{n=1}^\infty\frac{n}{(nk+r)^s(nk+r+1)^s}=\frac{2k}{(4\pi k)^s}\sum_{n=1}^\infty\frac{1}{n^{2s-1}}+\eta_k+\mathcal O(s-1),
\end{equation}
where
\begin{equation}
\begin{split}
\eta_k&\coloneqq\frac{1}{2\pi k}\sum_{r=0}^{k-1}\sum_{n=1}^\infty\left[\frac{nk^2}{(nk+r)(nk+r+1)}-\frac{1}{n}\right]\\
&=-\frac{1}{2\pi k}\sum_{r=0}^{k-1}\left[\gamma_{\rm E}-r\psi\left(1+\frac{r}{k}\right)+(1+r)\psi\left(1+\frac{r+1}{k}\right)\right]\\
&=-\frac{\gamma_{\rm E}+\psi(2)}{2\pi}=-\frac{1}{2\pi}
\end{split}
\end{equation}
where $\psi(z)\coloneqq\frac{\Gamma'(z)}{\Gamma(z)}$ is the digamma function and $\psi(2)=1-\gamma_{\rm E}$. On the other hand
\begin{equation}
\frac{2k}{(4\pi k)^s}\sum_{n=1}^\infty\frac{1}{n^{2s-1}}=\frac{1}{2\pi}\left[\frac{1}{2s-2}+\gamma_{\rm E}-\frac{\ln(4\pi k)}{2}+\mathcal O(s-1)\right]
\end{equation}
and 
\begin{multline}
\left(\frac{k}{4\pi}\right)^s\sum_{n=1}^\infty\frac{1}{(nk+r)^s(nk+r+1)^s}=\\
=\frac{1}{4\pi}\sum_{r=0}^{k-1}\left[\psi\left(1+\frac{r+1}{k}\right)-\psi\left(1+\frac{r}{k}\right)\right]+\mathcal O(s-1)
=\frac{1}{4\pi}+\mathcal O(s-1).
\end{multline}
Collecting all the pieces we obtain
\begin{equation}
Z(s)=\frac{1}{4\pi(s-1)}+\frac{k-2-\ln(4\pi k)}{4\pi}+\frac{\gamma_{\rm E}}{2\pi}+\mathcal O(s-1),
\end{equation}
which reduces to the case of the surface of the sphere if we put $k=1$.

Similar arguments can be repeated if Neumann boundary conditions are chosen. In this case, the eigenfunctions of the Laplacian are
\begin{equation}
\Psi_l^m(\theta,\phi)=\frac{Y_l^m(\theta,\phi)+iY_l^{-m}(\theta,\phi)}{\sqrt 2},\quad l,m\in\mathds N,
\end{equation}
with corresponding eigenvalue
\begin{equation}
\lambda_{m,l}=\frac{l(l+1)}{r^2},\quad l\in\mathds N_0,\quad 0\leq m\leq l,\ 2m=0\mod k.
\end{equation}
If $k=2\kappa$ is even, then we have to compute
\begin{multline}
Z(s)=\left(\frac{\kappa}{2\pi}\right)^s\left[\sum_{r=1}^{\kappa-1}\frac{1}{r^s(r+1)^s}+\sum_{r=0}^{\kappa-1}\sum_{n=1}^\infty\frac{n+1}{(n\kappa+r)^s(n\kappa+r+1)^s}\right]\\
=\frac{1}{4\pi(s-1)}+\frac{\kappa-1+\gamma_{\rm E}}{2\pi}-\frac{\ln(2\kappa\pi)}{4\pi}=\frac{1}{4\pi(s-1)}+\frac{k-2-\ln(k\pi)}{4\pi}+\frac{\gamma_{\rm E}}{2\pi}.
\end{multline}
If $k=2\kappa+1$, then $2m=0\mod k$ iff $m=0\mod k$: repeating the usual arguments, the same result is obtained, showing that the Kronecker's mass in the case of Neumann conditions differs from the periodic boundary conditions case by an overall $\frac{1}{2\pi}\ln 2$ constant.

%%%%%%%%%%%%%%%%%%%%%%%%%%%%%%%%%%%%%%%%%%%%%%%%%%%%%%%
\section{Kronecker's limit formulas}
\label{app:Kron}

\noindent
In this Appendix we will summarize some results obtained in the realm
of analytic number theory. Let $s\in \mathds{C}$. The Riemann $\zeta$-function
$\zeta(s)$ is defined in the half-plane $\Re(s)>0$ by
\begin{equation}
\zeta(s)\coloneqq\sum_{k\geq 1} \frac{1}{k^s}.
\end{equation}
The series converges absolutely for $\Re(s)\geq 1 + \epsilon$ for
every $\epsilon > 0$. Riemann proved that $\zeta(s)$ has an analytic
continuation in the whole $s$-plane which is regular except a simple
pole at $s=1$ with residue $1$. At $s=1$, $\zeta(s)$ has an expansion 
\begin{equation}
\zeta(s)=\frac{1}{s-1}+\gamma_{\mathrm E} +o(s-1).
\end{equation}
As generalization of the Riemann $\zeta$-function, we consider a
positive-definite binary quadratic form, in the real variables $u,v
\in \mathds{R}$
\begin{equation}
Q(u,v)\coloneqq a u^2 + 2 b uv + c v^2
\end{equation}
where $a, b , c\in \mathds{R}$, $a>0$ and $d\coloneqq ac-b^2>0$. Let us define
\begin{equation}
 \zeta_Q(s)\coloneqq\sum_{\substack{(m,n)\in \mathds{Z}^2\\ n^2+m^2\neq 0}}\frac{1}{[Q(m,n)]^{s}}.
\end{equation}
Now
\begin{equation}
 Q(u,v) = a \left( u + \frac{b}{a} \right)^2 + \frac{v^2d}{a} = a
 \left( u + \frac{b + i\sqrt d }{a} v \right)\left( u + \frac{b -
   i\sqrt d}{a} v \right)
 = a |u + \tau v|^2
\end{equation}
 where
 \begin{equation}
 \tau = \frac{b + i\sqrt d}{a}
 \quad\text{with }\Im(\tau)=a^{-1}\sqrt d>0.
 \end{equation} 
If $d=1$, $\zeta_Q(s)\equiv\zeta_\tau(s)$ given in Eq.~\eqref{Kro1repeat}, associated to $Q$, is
defined for $\Re(s)>1$   can be analytically continued into a regular
function for $\Re(s)>\sfrac{1}{2}$ except for a simple pole at $s=1$
with residue $\pi$, and the function $\zeta_Q(s)$, has an expansion
(first limit formula of Kronecker)
\begin{equation}
\label{Kro1}
  \zeta_\tau(s) =\frac{\pi}{s-1}+2 \pi \left[ \gamma_{\mathrm{E}} -
    \ln ( 2\sqrt{\Im(\tau)}|\eta(\tau)|^2 )\right] + o(s-1), 
\end{equation}
where
\begin{equation}
 \eta(s)\coloneqq \e^{\tfrac{\pi  i s}{12}}\prod_{n=1}^\infty\left(1-\e^{2\pi i n s}\right)
\end{equation}
is the Dedekind $\eta$-function, which satisfies the functional equations
\begin{subequations}
\begin{align}
\eta(s+1) = & \e^{\tfrac{\pi  i}{12}}\, \eta(s) \vphantom{\frac{1}{2}}\\
\eta\left(-\tfrac{1}{s}\right) = & \sqrt{-i s} \eta(s) \, .
\end{align}
\end{subequations}
Known particular values are
\begin{subequations}
\begin{align}
\eta(i)&=\frac{\Gamma\left(\sfrac{1}{4}\right)}{2\pi^{\sfrac{3}{4}}}\\
\eta(2i)&=\frac{\Gamma\left(\sfrac{1}{4}\right)}{2^{\sfrac{11}{8}}\pi^{\sfrac{3}{4}}} \\
\eta(4i)& = \left(-1+\sqrt{2}\right)^{\sfrac{1}{4}} \frac{\Gamma\left(\sfrac{1}{4}\right)}{2^{\sfrac{29}{16}}\pi^{\sfrac{3}{4}}} \, .
\end{align}
\end{subequations}
For a complete discussion of the results sketched here, see for example~\cite{Siegel1965}.

\bibliographystyle{unsrt}
\bibliography{AssignmentANDTsp.bib}
\end{document}